\documentclass[reprint, prb, superscriptaddress]{revtex4-2}
\usepackage{amsmath,amssymb}
\usepackage{stmaryrd}
\usepackage{array}
\usepackage{latexsym}
\usepackage{CJK}
\usepackage{graphicx}
\usepackage{indentfirst}
\usepackage{listings}
\usepackage{xcolor}
\usepackage{booktabs}
\usepackage{stmaryrd}
\usepackage{multirow}
\usepackage{verbatim}
\usepackage{makecell}
\usepackage{lineno,hyperref}
\usepackage{longtable}
\usepackage{hyperref}
\usepackage{upgreek}
\usepackage{mathtools}
\usepackage{comment}
\usepackage{array}

\newcolumntype{L}[1]{>{\raggedright\arraybackslash}p{#1}} % Left aligned
\newcolumntype{C}[1]{>{\centering\arraybackslash}p{#1}} % Center aligned
\newcolumntype{R}[1]{>{\raggedleft\arraybackslash}p{#1}} % Right aligned
\hypersetup{colorlinks,breaklinks,linkcolor=blue,urlcolor=blue,anchorcolor=blue,citecolor=blue}

\newcommand{\rmd}{\mathrm{d}}

\begin{document}

%\title{Towards understanding realistic grain growth: Microstructure topology and internal stress}
\title{Shear-Coupled Grain Growth Statistics}

\author{Caihao Qiu}
\affiliation{Department of Materials Science and Engineering, City University of Hong Kong,  Hong Kong SAR, China}
\affiliation{Department of Mechanical Engineering, The University of Hong Kong, Pokfulam, Hong Kong SAR, China}
%\affiliation{Department of Materials Science and Engineering, Northwestern University, Evanston, IL 60208, USA}
\author{David J. Srolovitz}
\affiliation{Department of Mechanical Engineering, The University of Hong Kong, Pokfulam, Hong Kong SAR, China}
\affiliation{Materials Innovation Institute for Life Sciences and Energy (MILES), The University of Hong Kong, Shenzhen, China}
\author{Gregory S. Rohrer}
\affiliation{Department of Materials Science and Engineering, Carnegie Mellon University, Pittsburgh, PA, USA}
\author{Jian Han}
\affiliation{Department of Materials Science and Engineering, City University of Hong Kong,  Hong Kong SAR, China}
\author{Marco Salvalaglio}
\email{marco.salvalaglio@tu-dresden.de}
\affiliation{Institute of Scientific Computing, TU Dresden, 01062 Dresden, Germany}
\affiliation{Dresden Center for Computational Materials Science, TU Dresden, 01062 Dresden, Germany}

\date{\today}

\begin{abstract}
Grain growth (GG), driven by grain boundary (GB) migration, is a fundamental mechanism of microstructural evolution in polycrystalline materials. GB migration is frequently accompanied by a relative shear displacement of grains meeting at GBs, a phenomenon known as shear coupling. This coupling induces internal stresses within the microstructure, which recent studies have shown to play a decisive role in dictating the evolution of microstructure and GG pathways. This work provides a detailed characterization of the statistical features of two-dimensional GG in the presence of GB shear coupling through continuum modeling of GB migration that incorporates fundamental microscopic mechanisms and diffuse-interface simulations. We demonstrate that incorporating shear coupling produces a more heterogeneous, less equiaxed microstructure than conventional curvature-driven GG, while yielding topological and geometric properties consistent with experimental and atomistic observations. We further demonstrate that as grain grows, internal stress relaxes. Highly stressed grains shrink faster, and lightly stressed grains grow faster than other grains. These findings demonstrate that internal stress, an intrinsic feature of GG, profoundly changes essential features of GG microstructure and kinetics, consistent with experiments and atomic-scale simulations.
\end{abstract}

\maketitle
\section{Introduction}

Polycrystals are the most common type of engineering material, composed  of crystalline domains (grains) with different crystallographic orientations separated by grain boundaries (GBs).
The size, morphology, and topology of grains critically influence macroscopic material properties. 
Among the various mechanisms of microstructure evolution in polycrystals, grain growth (GG) plays a particularly prominent role. 
It occurs during thermal and thermomechanical treatments over a wide range of temperatures and length scales and is often deliberately exploited to engineer the grain size distribution and crystallographic texture, or suppressed to stabilize fine-grained microstructures. 

The primary driving force for GG is capillarity.
Since GBs are crystal defects with a positive excess energy, they migrate, particularly at elevated temperatures, to reduce the total GB area. 
This driving force leads to GB migration toward their centers of mean curvature; this is mean curvature or capillarity-driven flow. 
Microstructure evolution during isotropic GG has been well characterized over the past several decades (e.g., see~\cite{mason2015geometric}).
Anisotropic GB energies and mobilities produce weighting factors on the curvature, amongst other effects such as faceting. 
The collective outcome of this GB migration is a progressive decrease in the number of grains and a corresponding increase in the average grain volume, as established by classical theories of normal GG~\cite{burke1952recrystallization,von1952metal,mullins1956two,macpherson2007vonneumann} (the effect of GB anisotropy on GG has also been examined~\cite{kim2024statistics,naghibzadeh2024impact}). 
A number of experiments, atomistic simulations, and analyses, however, point to additional effects which cannot be simply ascribed to anisotropic GB energy or mobilities, including stress-driven GG~\cite{legros2008situ,sharon2011stress,winning2006grain} and grain rotation during GG~\cite{harris1998grain,farkas2007linear,qiu2024disconnection,tian2024grain}, and effective mobilities poorly correlating with GB degrees of freedom~\cite{zhang2020grain,qiu2024grain}.

Grain boundary migration is often accompanied by shear translations of one grain relative to another across GBs, a phenomenon known as shear-coupled migration~\cite{cahn2006coupling}. 
%This is typically quantified via shear coupling factor $\beta = v_\parallel / v_\perp$, where $v_\parallel$ and $v_\perp$ are the rates of shear and migration.
This phenomenon is mechanistically described by the flow of line defects that glide along GBs and have step and/or dislocation character, namely disconnections~\cite{bollmann1970general,ashby1972boundary,hirth1973grain}. 
The step nature of disconnections describes GB curvature, and their motion leads to curvature flow.   
Moreover, disconnection motion generates internal stress and couples to external stress owing to their Burgers vectors~\cite{han2018grain}. 
Disconnections have been widely observed on GBs~\cite{rajabzadeh2013evidence,zhu2019situ} and heterophase interfaces~\cite{pond2003comparison,zheng2018determination,qi2020interdiffusion}. 
The generation of internal stresses within the microstructure during GG has been observed in experiments~\cite{gautier2025quantifying,tian2024grain} and atomistic simulations~\cite{thomas2017reconciling,guo2013generalized}.

Phenomenological models have been developed to incorporate the effects of mechanical stress into GB migration and to assess its impact on GG. 
In one class of approaches, GBs are modeled in terms of geometrically necessary dislocations (GNDs), which can describe grain rotation and stress-driven GG~\cite{admal2018unified,he2021polycrystal}. 
Other formulations explicitly account for dislocation densities stored within the grains, such that differences in the associated elastic energy across a GB act as an additional thermodynamic driving force for GB migration~\cite{huang2024phase}. 
Elastic energy contributions associated with shear-coupled GB migration have also been introduced by effectively assigning a shear-coupling factor to individual grains, resulting in a shear-coupling ``jump'' at the boundary~\cite{gokuli2021multiphase} and have been included in reduced-order modeling of microstructure evolution~\cite{bugas2024grain}. 
We recently extended a bicrystallography-respecting description of disconnection-mediated GB migration to microstructure evolution~\cite{han2022disconnection,sal2022disconnection}. 
This framework uniquely enables the study of relatively large microstructures consistent with observed microscopic mechanisms of GB migration.
This work demonstrated that GB shear coupling and the resulting internally generated stresses are essential ingredients for a realistic description of GG, in quantitative agreement with both experimental observations and atomistic simulations~\cite{qiu2025why}.

Despite this significant progress, a comprehensive understanding of how internal and external stresses shape GG and its microstructures, as well as how internal stresses evolve in space and time, remains lacking. 
Large-scale characterization of GG has been effectively described through statistical analyses of the geometric and topological properties of the microstructure~\cite{dehoff1975topography}. 
Such characterizations have been extensively reported from experimental measurements~\cite{kurtz1980microstructure, fradkov1985experimental,barmak2013grain}. 
On the theoretical and numerical side, capillarity-driven microstructure evolution has been investigated using Monte Carlo Potts model~\cite{srolovitz1984computer,grest1988domain,mulheran1991simple}, phase-field methods~\cite{fan1997computer,krill2002computer,kamachali20123}, and front-tracking simulations~\cite{lazar2010more,lazar2011more,mason2015geometric,lazar2020distribution}. 
Nevertheless, systematic discrepancies persist between microstructure statistics obtained from experiments or atomistic simulations and those predicted by classical curvature-driven models~\cite{krill2002computer,barmak2013grain,backofen2014capturing,la2019statistics}, pointing to missing physical ingredients that are required for a  realistic description of GG.

In this work, we aim to close this gap by employing GG simulations that incorporate shear-coupled GB migration ~\cite{sal2022disconnection,qiu2025why}. 
We investigate the role of internal stress and different types of external stress on the GG and the statistics of GG microstructure topology and geometry. 
The internal stress field is studied both by accounting for its spatial variation and through averaging within grains. 
We further explore the correlation between the evolution of internal stress and the evolution of various geometric properties. 
These results provide a new understanding of the essential role of shear coupling and stresses in the evolution of polycrystalline microstructures, which may be useful for controlling microstructure evolution and optimizing microstructures.

%%%% FIG GRAIN GROWTH
\begin{figure*}[t!]
\includegraphics[width=0.9\linewidth]{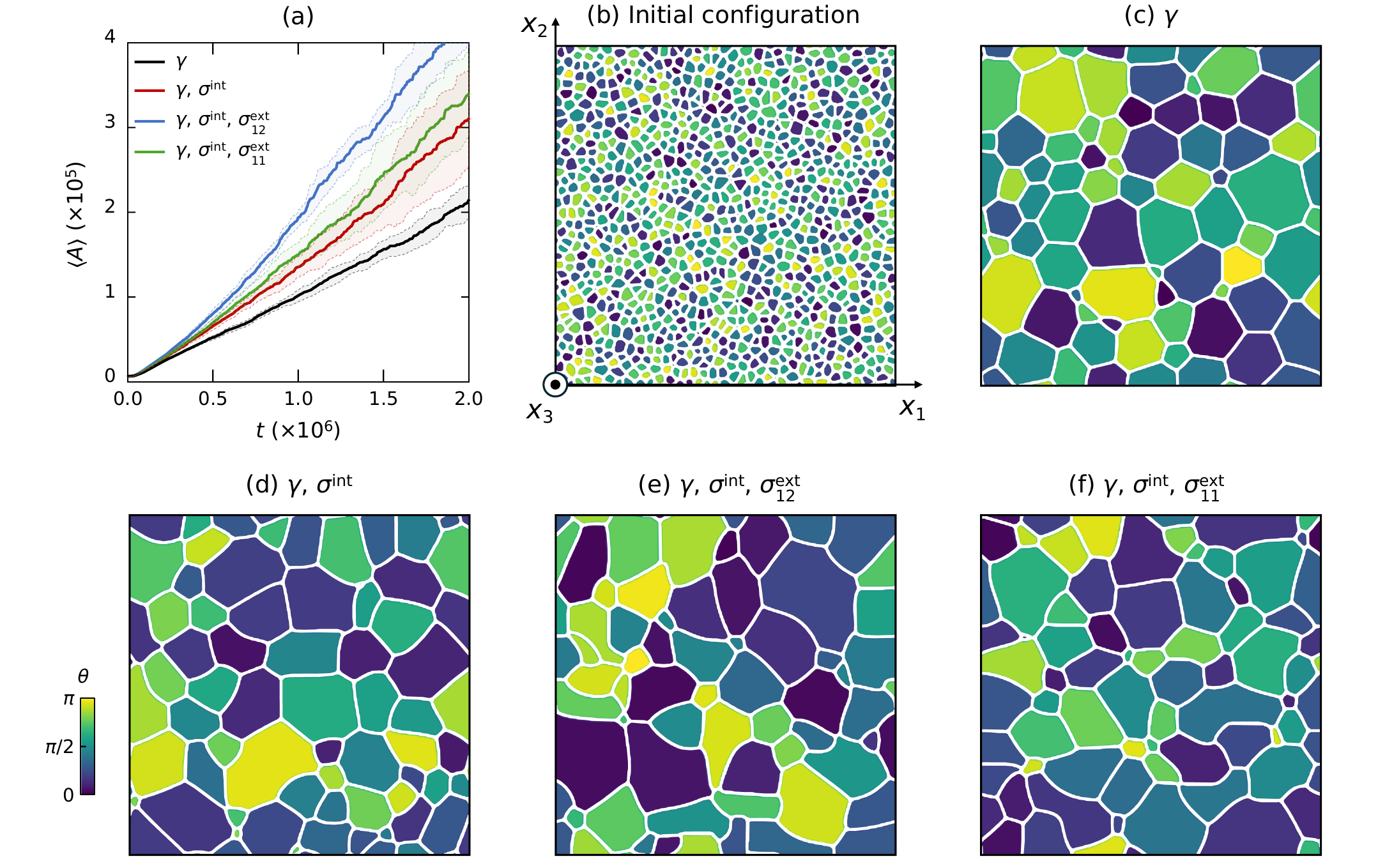}\hspace{-1.78em}%
\caption{(a) Mean grain size (area) $\langle A \rangle$ vs. time $t$ during grain growth. 
Each curve is averaged over 10 independent PF simulations with and without internal stress ($\sigma^{\rm int}$) and applied external stress ($\sigma^{\rm ext}$).
The shaded regions indicate the standard deviation from the 10 initially random microstructures. 
(b) Initial microstructure (with a white GB network and colored orientation $\theta$ field) of a typical 1000-grain polycrystal. 
All microstructures are shown at the same mean grain area $\langle A \rangle \approx 1\times10^5$, which evolved from the initial microstructure in the cases
(c) of mean curvature flow ($\gamma$), (d) with $\sigma^{\rm int}$ and no $\sigma^{\rm ext}$, (e) with $\sigma^{\rm int}$ and applied external shear stress $\sigma^{\rm ext}_{\rm 12}$, and (f) with $\sigma^{\rm int}$ and external tensile stress $\sigma^{\rm ext}_{\rm 11}$. 
{\color{red}{}}
}
\label{fig_law}
\end{figure*}

%%%% TABLE 1
\begin{table*}[t]\label{tab:parameter}
\begin{center}
\caption{Dimensionless parameters used in the simulations.}
\renewcommand{\arraystretch}{1.5}
%\begin{ruledtabular}
\small
%\begin{tabular}{ m{4cm}<{\centering} m{2.8cm}<{\centering} m{2.8cm}<{\centering} m{2.8cm}<{\centering} m{3cm}<{\centering}}
\begin{tabular}{ ccccc}
\hline\hline
~~~Case~~~~                                                        & ~~~GB energy, $\gamma$~~~ & ~~~GB mobility, $M$~~~              & ~~~Shear modulus, $\mu$~~~ & ~~~Applied external stress, $\sigma^{\rm ext}$~~~                \\ \hline
$\gamma$ (mean curvature flow)               & 1                   & $\pi^2/(2\varepsilon)\approx 0.1$ & 0                    & 0                                      \\ %\hline
$\gamma,~\sigma^{\rm int}$                                      & 1                   & $\pi^2/(2\varepsilon)\approx 0.1$ & 6                    & 0                                      \\ %\hline
$\gamma,~\sigma^{\rm int},~\sigma^{\rm ext}_{\rm 12}$ & 1                   & $\pi^2/(2\varepsilon)\approx 0.1$ & 6                    & $\sigma^{\rm ext}_{\rm 12} = 0.3$ \\ %\hline
$\gamma,~\sigma^{\rm int},~\sigma^{\rm ext}_{\rm 11}$   & 1                   & $\pi^2/(2\varepsilon)\approx 0.1$ & 6                    & $\sigma^{\rm ext}_{\rm 11} = 0.3$   \\ \hline\hline
\end{tabular}
%\end{ruledtabular}
\end{center}
\end{table*}

%%%%%%%%%%%%%%%%%%%%%%%%%%%%%%%%%%%%%%%%%%%%%%%%%%%%%%%%%%%%%%%%%%%
\section{Methods}

We exploit a multi-phase field (PF) model reproducing GB migration as described by the equation of motion in the presence of disconnection flow~\cite{zhong2019grain,han2022disconnection}, as proposed in our earlier studies~\cite{sal2022disconnection,qiu2025why}. 
In brief, the polycrystalline microstructure is characterized by a set of order parameter fields $\boldsymbol{\eta}(\mathbf{x}) = \{\eta_1(\mathbf{x}), \eta_2(\mathbf{x}), \cdots, \eta_i(\mathbf{x}), \cdots, \eta_{N_g}(\mathbf{x}) \}$, with $\mathbf{x} \in \Omega \subset \mathbb{R}^2$ and $N_g$ is the total number of grains.
The order parameters $\eta_i$ represent the local volume fraction of grain $i$; hence, a local constraint must be imposed~\cite{Steinbach1999,Steinbach_2009}, namely, $\sum_i^{N_g} \eta_i(\mathbf{x}) = 1$.
%\begin{equation}\label{eq:constraint}
%    \sum_i^{N} \eta_i(\mathbf{x}) = 1. 
%\end{equation}
Inside grain $i$, $\eta_i = 1$ and $\eta_j = 0$ for $i \neq j$.
At the GB between grains $i$ and $j$, $\eta_i \eta_j > 0$, $\eta_k = 0$ for $k\neq i\neq j$. At the TJ among grains $i$, $j$ and $k$, $\eta_i \eta_j \eta_k > 0$, $\eta_h = 0$ for $h \neq i \neq j \neq k$.

The evolution of the order parameter field $\eta_i$ incorporating the constraint %\eqref{eq:constraint}, 
is given by
\begin{equation}\label{eq:pf-dynamic}
\begin{split}
    \frac{\partial \eta_i}{\partial t} &= \sum_j^{N_g} M_{ij}\left(f_{\rm GB} + f_{\rm bulk} + f_{\rm elastic}\right),\\
f_{\rm GB} &=\gamma_{ij} \left[\eta_j \nabla^2\eta_i - \eta_i \nabla^2\eta_j - \frac{\pi^2}{2\varepsilon^2}\left(\eta_j - \eta_i\right)  \right], \\
f_{\rm bulk} &= \frac{\pi}{\varepsilon} \sqrt{\eta_i\eta_j} \left(\psi_j - \psi_i\right), \\
f_{\rm elastic} &= \boldsymbol{\tau}\cdot\boldsymbol{\beta}_{(i,j)}|\nabla\eta_j|\left[\hat{\mathbf{n}}(\eta_i) \cdot \hat{\mathbf{n}}(\eta_j)   \right],
\end{split}
\end{equation}
with $M_{ij}$ and $\gamma_{ij}$ refer to the mobility and GB energy density of GB-$ij$, respectively, $\varepsilon$ is a parameter that relates to the GB thickness as realized in the considered diffuse domain approach, $\psi_i$ and $\psi_j$ are the bulk free energy densities of grains $i$ and $j$. 
$\boldsymbol{\tau} =  \boldsymbol{\tau}^{\rm int} + \boldsymbol{\tau}^{\rm ext}$ encodes the shear stress across the GB with $\boldsymbol{\tau}\equiv(\tau^{(k)},\tau^{(k+1)})$, while $k$ and $k+1$ are index reference low-energy GBs (chosen from a prescribed set) which are the closest in orientation to GB-$ij$~\cite{han2022disconnection}.
$\boldsymbol{\tau}$ results from the superposition of internal stresses (int) and external applied stresses (ext), and $\boldsymbol{\beta}_{(i,j)}\equiv(\beta_{(i,j)}^{(k)},\beta_{(i,j)}^{(k+1)})$ are the corresponding GB shear coupling factors. 
Here, $\boldsymbol{\tau}^{\rm ext}$ is constant and
$\boldsymbol{\tau}^{\rm int}(\mathbf{x})$  is obtained by extending the corresponding quantities for the underlying sharp interface equation~\cite{han2022disconnection} in the direction normal to the GB (as in Level Set approaches~\cite{sethian1999level}), $\tilde{\boldsymbol{\tau}}^{\rm int}(\mathbf{s})$ with $\mathbf{s} \in \Gamma\equiv \bigcup_{ij} \Gamma_{ij}$ and $\Gamma_{ij}$ the nominal sharp-interface between grains $i$ and $j$. 
The internal shear stresses are computed from the stress tensor 
\begin{equation}
    \tilde{\sigma}_{\ell m}^{\rm int}(\bar{\mathbf{s}}) = K\sum_i^{N_g}\sum_j^{N_g} \int_\Gamma \boldsymbol{\beta}_{(i,j)} (\mathbf{s}) \cdot \mathbf{l}_{(i,j)} (\mathbf{s})\tilde{\sigma}_{\ell m}^{\rm edge}(\bar{\mathbf{s}} - \mathbf{s}) \rmd \mathbf{s}, 
\end{equation}
where $\boldsymbol{\beta}_{(i,j)}$ is the shear coupling factor for a segment of GB between grains $i$ and $j$, $K = \mu/[2\pi(1-\nu)]$, $\mu$ is the shear modulus, $\nu$ is the Poisson's ratio, $\mathbf{l}_{(i,j)}$ is local tangent vector of the GB, ${\sigma}_{\ell m}^{\rm edge}$ is the stress field component of an edge dislocation located at the origin~\cite{anderson2017} with Burgers vector dictated by $k^\text{th}$ reference GB. 

The 2D polycrystalline microstructures in our simulations are assumed to be a cross-section through a $\langle 110 \rangle$ columnar microstructure of a face-centered cubic (FCC) material. 
%^We label the $[001]$\parallel\mathbf{x_1}$, $[1\bar{1}\bar{2}]\parallel\mathbf{x_2}$ and $[110]\parallel\mathbf{x_3}$. 
For this polycrystal,  GBs can be described in terms of an inclination between two symmetric tilt GBs (with normals $[001]$ and $[1\bar{1}\bar{2}]$); i.e.,  a two-reference system description~\cite{han2022disconnection}. 
The local GB shear coupling factor vector $\boldsymbol{\beta}_{(i,j)}$ is determined by the GB misorientation angle $\Delta\theta_{ij}=\theta_i-\theta_j$  (the inclination dependence is automatically accounted for in the two-reference system description) according to~\cite{han2018grain}:
\begin{equation}\label{beta_microstructure}
\begin{aligned}
    &\beta_{(i,j)}^{(1)}(\Delta\theta_{ij}) =
    S\left\{
        \begin{aligned}
        &2\tan\left(\dfrac{\Delta\theta_{ij}}{2}\right),~ 0\leq\Delta\theta_{ij} < \alpha_1 \\
        &2\tan\left(\dfrac{\Delta\theta_{ij} - {\alpha_0}}{2}\right),~ \alpha_1 \leq \Delta\theta_{ij} < \alpha_2  \\
        &2\tan\left(\dfrac{\Delta\theta_{ij} - \pi}{2}\right),~\alpha_2 \leq \Delta\theta_{ij} < \pi
        \end{aligned}
        \right. \\
         &\beta_{(i,j)}^{(2)}(\Delta\theta_{ij}) = \beta_{(i,j)}^{(1)}(\pi - \Delta\theta_{ij}),
\end{aligned}
\end{equation}
where $\beta^{(1)}$ and $\beta^{(2)}$ are the shear coupling factors of the two symmetric reference GBs, $\alpha_0=2\tan^{-1}(\sqrt{2})$ (i.e., the misorientation angle of an incoherent twin GB in FCC), $\alpha_1 = 7\pi/18$, $\alpha_2 = 5\pi/6$, and $S$ is a scaling factor (see below). 

The  parameters used in this work are summarized in Table.~\ref{tab:parameter}. { Note, the dimensionless parameters employed in the simulations are approximately those of aluminum. However, we explicitly assume that  GB energies are isotropic in the current study to investigate how shear-coupled GB migration alters conventional mean curvature flow (i.e., the von Neumann-Mullins relation).} We set $S=0.033$ to account for the observation that shear coupling factors are smaller in polycrystals relative to bicrystals (by roughly this amount, associated with mechanical constraint); see~\cite{rajabzadeh2013evidence,thomas2017reconciling,gautier2025quantifying}.

We employ the following non-dimensionalisation of variables in the simulations: (1) lengths $\tilde{\mathbf{x}} = \mathbf{x}/a_0$, (2) time intervals $\Delta\tilde{t} = \Delta t M_0\gamma_0/a_0^2$, (3) GB mobility $\tilde{M}_{ij} = M_{ij}/M_0$, (4) stress $\tilde{\sigma} = \sigma a_0/\gamma_0$, (5) bulk free energy density $\tilde{\psi} = \psi a_0/\gamma_0$, (6) GB energy $\tilde{\gamma} = \gamma/\gamma_0$, where 
$a_0=1$ nm  ~and $\gamma_0=1$ J/m$^2$.
The simulation cell is chosen to be square and is discretized into 250$\times$250 grid points with uniform discretization $\delta x_1 =\delta x_2 = \varepsilon/5 = 10$. 
Numerical simulations are performed by integrating Eq.~\eqref{eq:pf-dynamic} with time steps $\Delta t = 10$ via a simple finite-difference scheme.

The initial microstructures for PF simulations are Voronoi tessellations of 1000 randomly distributed points (Poisson-distributed); see an example in Fig.~\ref{fig_law}b. 
For each investigation, we generate 10 independent initial microstructures with randomly distributed GB networks and grain orientations; e.g., see Fig.~\ref{fig_law}b.
The results reported below are averages over the evolution of  10 random initial samples.

%%%%%%%%%%%%%%%%%%%%%%%%%%%%%%%%%%%%%%%%%%%%%%%%%%%%%%%%%%%%%%%%%%%

\begin{table}[t]\label{tab:fit_parameter_size}
\centering
\caption{Fitting parameters of the mean grain area vs. time data in the form $\langle A \rangle = K_A t^{n_A}$.}
\renewcommand{\arraystretch}{1.5}
%\begin{ruledtabular}
\small
%\begin{tabular}{ m{4cm}<{\centering} m{2.8cm}<{\centering} m{2.8cm}<{\centering} m{2.8cm}<{\centering} m{3cm}<{\centering}}
\begin{tabular}{ccc}%{p{3cm} p{2.5cm} p{2.5cm}}
\hline\hline
\quad\quad\quad Case\quad\quad\quad & \quad\quad\quad\quad $K_A$ \quad\quad\quad\quad              &  \quad\quad\quad\quad $n_A$ \quad\quad\quad\quad   \\ \hline
$\gamma$                                      & 9.460 $\pm$ 0.2473                   & 1.006 $\pm$ 0.0026                                     \\ %\hline
$\gamma,~\sigma^{\rm int}$                                      & 4.429 $\pm$ 0.0729                   & 1.125 $\pm$ 0.0017                                     \\ %\hline
$\gamma,~\sigma^{\rm int},~\sigma^{\rm ext}_{\rm 12}$   & 3.022 $\pm$ 0.0941                   & 1.201 $\pm$ 0.0032  \\ %\hline
$\gamma,~\sigma^{\rm int},~\sigma^{\rm ext}_{\rm 11}$   & 3.565 $\pm$ 0.0924                   & 1.157 $\pm$ 0.0027   \\ \hline\hline
\end{tabular}
%\end{ruledtabular}
\end{table} 

\section{Results}

\subsection{Grain size evolution}\label{sec:law}

We first illustrate the impact of internal stress and applied external stress on  GG; namely, the scaling of grain size with time in four different scenarios: (i) GG driven by curvature flow without internal stress ($\gamma$), (ii) GG driven by curvature flow and internal stress ($\gamma,~\sigma^{\rm int}$), (iii) GG driven by curvature flow, internal and applied external shear stress ($\gamma,~\sigma^{\rm int},~\sigma^{\rm ext}_{\rm 12}$), and (iv) driven by curvature flow, internal and applied external tensile stress ($\gamma,~\sigma^{\rm int},~\sigma^{\rm ext}_{\rm 11}$).
Figure~\ref{fig_law} shows the evolution of the mean grain area $\langle A \rangle$ during GG, along with the evolving microstructures, for the four cases.

\begin{figure}[t]
\includegraphics[width=0.9 \linewidth]{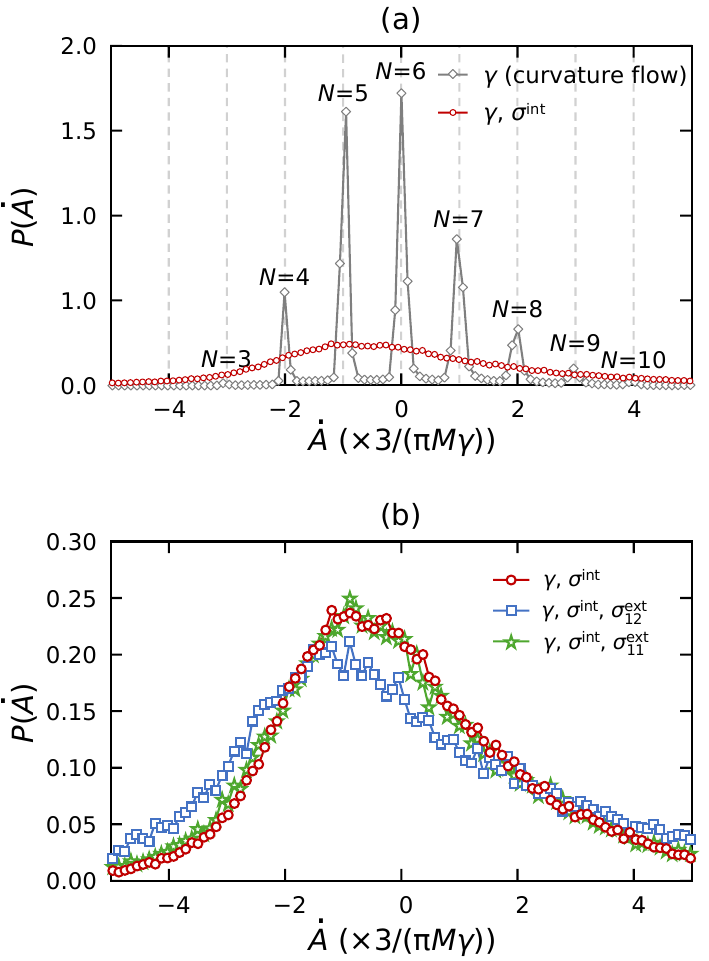}\hspace{-1.78em}%
\caption{Probability distribution of growth rates of individual grains in the cases (a) of curvature flow $\gamma$ (grey diamond) and with internal stress $\gamma,~\sigma^{\rm int}$ (red circles), and with additional applied external (b) shear stress $\gamma,~\sigma^{\rm int},~\sigma^{\rm ext}_{12}$ (blue squares) and tensile stress $\gamma,~\sigma^{\rm int},~\sigma^{\rm ext}_{11}$ (green star).
}
\label{fig_neumann}
\end{figure}
 
Figure~\ref{fig_law}a shows that including internal stresses accelerates GG, and applying external (both shear and tensile) stresses accelerates it further.
This is consistent with the experimental observation of stress-driven GG ~\cite{legros2008situ,sharon2011stress}.

The evolution of mean grain size is commonly fitted to a power law $\langle A \rangle = K_A t^{n_A}$ ($K_A$ is a constant and $n_A$ is known as the GG exponent); the best fit parameters for our data are shown in Table~\ref{tab:fit_parameter_size}.
Normal GG (i.e., mean curvature flow) in an isotropic system leads to a GG exponent $n_A=1$, in agreement with conventional GG theory~\cite{burke1952recrystallization}.
However, when internal stresses, generated during GG, are included in the simulations, the GG exponent increases to $n_A\approx1.125$; i.e., the rate of GG increases with time (we return to this observation in the Discussion).
Application of an external (shear or tensile) stress further increases the GG exponent; this effect is more pronounced for applied shear than tension. 
Interestingly, while application of an external stress increases the GG exponent $n_A$, it lowers the GG prefactor $K_A$ by an even greater factor.
This is consistent with the well-known compensation effect in GG~\cite{gottstein1998compensation}.

In the 1950s, von Neumann and Mullins ~\cite{von1952metal,mullins1956two} showed that mean curvature flow leads to the rate of growth of any grain (of area $A$) in a polycrystal described by
\begin{equation}
    \dot{A} \equiv \frac{\rmd A}{\rmd t} =  -2\pi M\gamma \left(1-\frac{1}{6}N\right),
\end{equation}
where $N$ is the number of neighbors/sides of grains (this has been extended to all dimensions~\cite{macpherson2007vonneumann}). 
Hence, the distribution of GG rates during isotropic mean curvature flow $P(\dot{A})$ is discrete. 
This is consistent with our curvature flow PF simulations (see Fig.~\ref{fig_neumann}a) with minor broadening associated with the diffuse interface nature of this simulation method.
However, when internal stresses are included in the simulation, the GG rate distribution is completely different, as seen in Fig.~\ref{fig_neumann}a. 
That is, with internal stresses, the growth rate distribution is completely smooth, with no remnants of the delta-function-like distribution associated with mean curvature flow and the classic von Neumann–Mullins relation. 
The smooth form of the GG rate distribution survives the application of external stresses, as seen in Fig.~\ref{fig_law}b.
Interestingly, the application of a shear stress shifts the GG rate distribution toward smaller growth rates, while the application of a tensile stress leaves it unchanged relative to curvature flow alone (Fig.~\ref{fig_law}b).

\begin{figure}[t]
\includegraphics[width=0.9\linewidth]{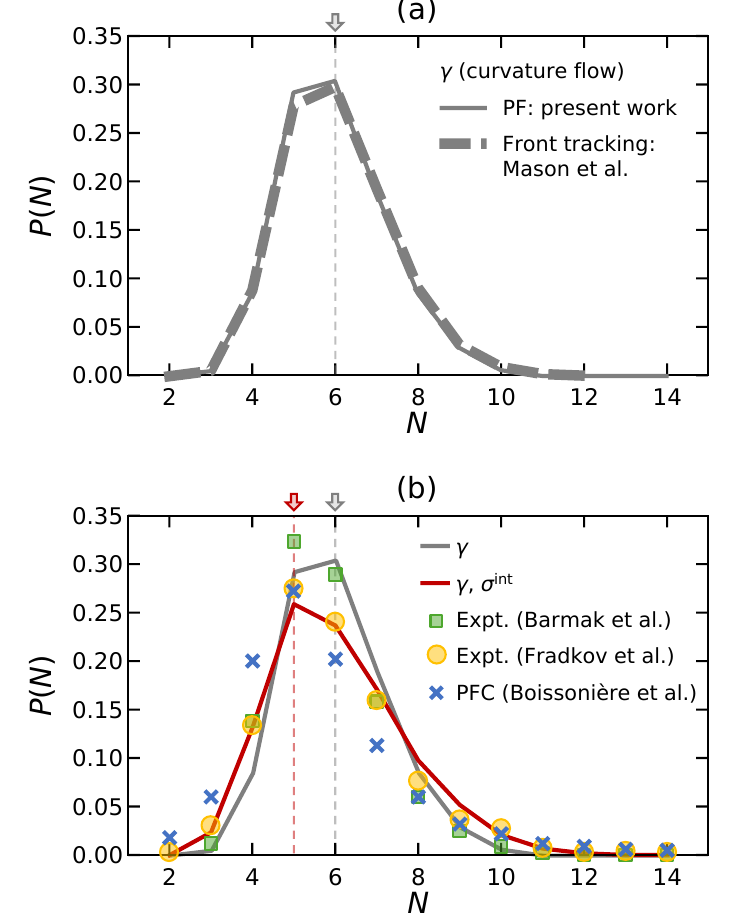}\hspace{-1.78em}%
\caption{Probability distributions of number of neighbors ($N$) of individual grains during GG described by (a) $\gamma$ (curvature flow), where the solid and dashed lines represent the results from the present  and previous front-tracking modeling work~\cite{mason2015geometric}, respectively; and (b) $\gamma, \sigma^{\rm int}$, where the orange circles, green squares, blue crosses, and the red line are the results from GG experiments in an Al thin film~\cite{fradkov1985experimental,barmak2013grain}, PFC simulation~\cite{la2019statistics} and the present work, respectively.
}
\label{fig_N}
\end{figure}

\subsection{Microstructural topology and geometry}

Figures~\ref{fig_law}c–f show microstructures with a mean grain size of $\langle A \rangle \approx 1\times 10^5$ that evolved from the same initial configuration in Fig.~\ref{fig_law}b under the four GG scenarios. 
While all of the microstructures appear very similar (i.e., GBs meet at $120^\circ$, 6 sides per grain on average, $\cdots$), there are also notable differences; e.g., the GBs appear more curved, and grains are less equiaxed when stress effects are included.
In this section, we characterize the effects of internal and external stresses on the geometrical and topological features of evolved microstructures.

The topological characteristics of a 2D microstructure are commonly described by the probability distribution of the number of grain neighbors (or sides), $P(N)$. 
In PF simulations, $P(N)$ peaks at $N = 6$ when GG is driven solely by curvature flow (see the solid line in Fig.~\ref{fig_N}a), consistent with previous independent numerical studies, such as the front-tracking (sharp-interface) results shown by the thick dashed line in Fig.~\ref{fig_N}a~\cite{mason2015geometric}. 
(Note, the front tracking simulations are based on a thousand times as many grains as our PF simulations, so they should be viewed as more accurate than our PF results - nonetheless, the agreement is excellent.)  
When GB shear coupling and internal stress are included, the peak of $P(N)$ shifts from $N = 6$ to $N = 5$ (see Fig.~\ref{fig_N}b). 
Comparison of the PF GG simulations results with experiment~\cite{barmak2013grain,fradkov1985experimental} and diffusive-timescale ``atomistic'' (phase field crystal, PFC) simulations~\cite{la2019statistics} shows very good agreement only when internal stresses are included.    %trend agrees with grain-growth experiments in thin films (green squares~\cite{barmak2013grain} and orange circles~\cite{fradkov1985experimental}) as well as with 2D phase-field-crystal (PFC) simulations (blue crosses~\cite{la2019statistics}). 
In particular, note that the peaks in the side distributions in the experiments, atomistic simulations, and PF with internal stress simulation are all at 5 (Fig.~\ref{fig_N}) rather than 6 (see Fig.~\ref{fig_N}a for two sets of curvature flow simulations).
%(Notably, our computed $P(N)$ closely matches the experimental distribution reported by Fradkov et al. for aluminum thin films (orange circles)~\cite{fradkov1985experimental}.

\begin{figure*}[t!]
\includegraphics[width=0.9995\linewidth]{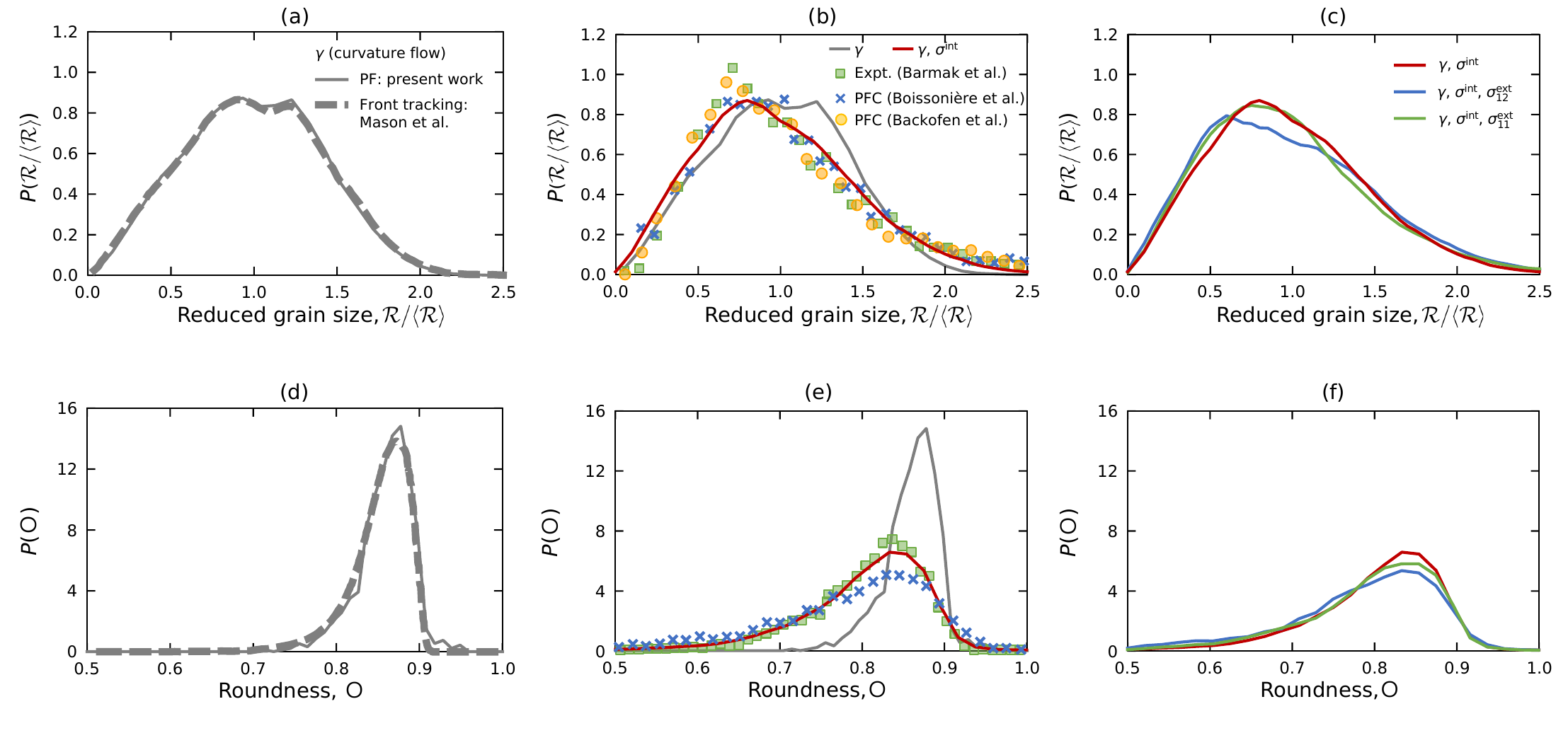}\hspace{-1.78em}%
\caption{Geometric properties of microstructures. probability distributions of (a)-(c) reduced grain size $\mathcal{R}/\langle \mathcal{R} \rangle$ and (d)-(f) grain roundness ${\rm \bigcirc}$. 
The grey, red, green, and blue lines correspond to our PF results for the case of curvature flow with internal stress, external tensile stress, and external shear stress, respectively.
The thick grey dotted lines in (a) and (d) are the results obtained by the front-tracking method~\cite{mason2015geometric}.
Green squares, blue crosses, and yellow circles are the results from experiments in an Al thin film~\cite{barmak2013grain} and PFC simulations~\cite{la2019statistics,backofen2014capturing}.
Note that each probability distribution is the averaged result based on microstructures with a mean grain area $\langle A \rangle > 5\times 10^4$.
}
\label{fig_size}
\end{figure*}

Another common descriptor of the microstructure is the grain size distribution (probability distribution of the reduced grain size), $P(\mathcal{R}/\langle \mathcal{R} \rangle)$ (here, $\mathcal{R} = \sqrt{A}$).
For isotropic curvature-driven GG, $P(\mathcal{R}/\langle \mathcal{R} \rangle)$ exhibits two peaks, at $\mathcal{R}/\langle \mathcal{R} \rangle \approx 0.9$ and $1.2$ (observed both in our PF simulations and in  front-tracking simulations~\cite{mason2015geometric}); see Fig.~\ref{fig_size}a.
When GB shear coupling and the resulting internal stress are included, these two peaks merge and shift towards a smaller grain size,  $\mathcal{R}/\langle \mathcal{R} \rangle \approx 0.8$, as shown by the red curve in Fig.~\ref{fig_size}b.
Experiments and PFC simulations also display single-peak distributions, with peak locations in the range $0.7$–$0.8$, cf. the red curve and the discrete symbols (circles~\cite{backofen2014capturing}, squares~\cite{barmak2013grain}, and crosses~\cite{la2019statistics}) in Fig.~\ref{fig_size}b.
When an external stress is applied, the grain size distribution retains the single peak; see Fig.~\ref{fig_size}c. 
An applied shear has a much larger effect on the grain size distribution than the same magnitude tensile stress (Fig.~\ref{fig_size}), shifting the grain size distribution towards smaller grain size. 

We also examine several other descriptors of the grain geometry; i.e., the probability distribution of the grain roundness $P(\bigcirc)$ where roundness is defined as $\bigcirc = 4\pi A/\mathcal{C}^2$ (Figs.~\ref{fig_size}d–f), the reduced grain perimeter $P(\mathcal{C}/\langle \mathcal{C} \rangle)$ (Figs.~S4a–c in supplementary material, SM), and the reduced GB length distribution $P(\mathcal{L}/\langle \mathcal{L} \rangle)$ (Figs.~S4d–f in SM).
All linear measures of grain size ($\mathcal{R}$, $\mathcal{C}$, $\mathcal{L}$) show the same trends with respect to the comparison of the curvature flow results and the inclusion of internal (and external) stresses (cf. Figs.~\ref{fig_size} and SM Fig.~S4).
The roundness is a measure of compactness of the grain shape (i.e., how close the grain  is to a circle). 
As seen in Fig.~\ref{fig_size}e, the grain roundness distribution for isotropic curvature flow has a relatively sharp single peak with a maximum near $\bigcirc=0.88$; this distribution broadens and shifts to smaller roundness (peak at $\bigcirc=0.84$ when internal (and external) stresses are included in the PF simulations.  
Overall, we find that GG microstructure becomes more inhomogeneous in the presence of shear coupling.

\begin{figure}[t]
\includegraphics[width=0.825\linewidth]{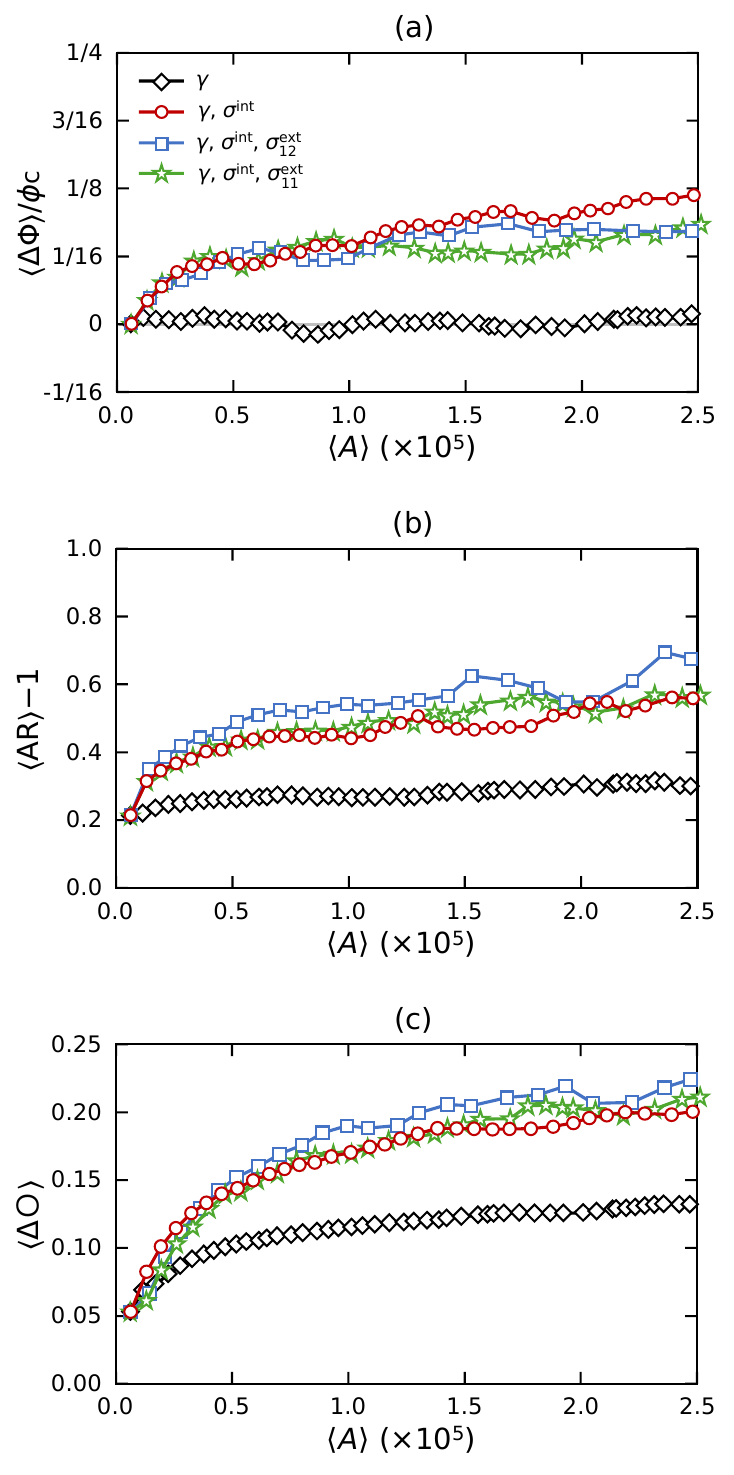}\hspace{-1.78em}%
\caption{Evolution of the average (a) GB degree of faceting $\langle\Delta\Phi\rangle/\phi_{\rm c}$, (b) grain aspect ratio $\langle{\rm AR}\rangle-1$ and (c) deviation from a circle $\langle \Delta \bigcirc\rangle$ vs. mean grain area in the cases of $\gamma$ (curvature flow, black circles); $\gamma,~\sigma^{\rm int}$ (red circles); $\gamma,~\sigma^{\rm int},~\sigma^{\rm ext}_{11}$ (green star); and $\gamma,~\sigma^{\rm int},~\sigma^{\rm ext}_{12}$ (blue squares).
}
\label{fig_nonsimilar}
\end{figure}

To quantify the degree of inhomogeneity, we examine several geometric properties of the simulated microstructure vs. mean grain size (or, equivalently, time - Fig.~\ref{fig_law}a); see Fig.~\ref{fig_nonsimilar}. 
To describe the degree of GB faceting in the microstructure, we first calculate the degree of faceting of each GB as $\Delta\Phi/\phi_{\rm c}$~\cite{qiu2023interface}
\begin{equation}
\Delta \Phi = \phi_{\rm c} - \int \min_k \left|\phi(s)-\phi^{(k)}(s)\right| \, \rmd s \, \Bigg/ \, \int \rmd s,
\end{equation}
where the integral over $s$ is along an entire GB, $\phi(s)=\arccos(\hat{\mathbf{l}}(s)\cdot \mathbf{e}_1)$ is the local inclination angle ($\hat{\mathbf{l}}(s)$ is the local tangent vector and $\mathbf{e}_1$ is the unit vector of $\mathbf{x}_1$ axis), and $\phi^{(k)} = 0$ for $k=0$ and $\phi^{(k)} = \pi/2$ for $k=1$.
$\phi_{\rm c}$ is the critical inclination angle when the GB is perfectly faceted, i.e., GBs are fully faceted when $\Delta\Phi/\phi_{\rm c} = 1$, whereas $\Delta\Phi/\phi_{\rm c} = 0$ corresponds to the absence of faceting. 
It is important to note that GB facets correspond roughly to the  reference interfaces for shear coupling. Note that for GBs with any anisotropy, these reference interfaces are highly coherent, relatively low-energy GB planes (corresponding to  cusps in the GB energy versus inclination). 
Even though the GB energies are isotropic, the GB shear coupling factors are anisotropic (inclination-dependent). 
This type of shear coupling-induced faceting need not be consistent with that from anisotropic GB energies (i.e., the  Wulff construction)~\cite{qiu2023interface}.
This is consistent with the observed lack of consistency between facet orientations and GB energy anisotropy in experiment~\cite{radetic2012mechanism}.

As seen in Fig.~\ref{fig_nonsimilar}a, the average degree of faceting $\langle\Delta\Phi\rangle/\phi_{\rm c}$ remains near constant under curvature flow conditions.
When internal stress and/or applied external stresses are included, GBs rapidly begin to facet, and the degree of faceting tends to increase over time.
Since external stress drives GB motion in a direction that does not necessarily favor faceting, the degree of faceting under external stress (blue squares and green star) is slightly smaller than that obtained when only internal stress is considered (red circles).

We also investigate the the average grain aspect ratio  $\langle {\rm AR} \rangle$ (the best fit of each grain shape to an ellipse  is determined, and the aspect ratio is the ratio of the major and minor axes); the deviation of grain aspect ratio from perfect equaixed morphology ($\langle {\rm AR} \rangle-1$) is shown in Fig.~\ref{fig_nonsimilar}b.
Similar to the GB degree of faceting, the aspect ratio appears to approach a steady state in the case of curvature flow.
However, when internal stresses are included in the microstructure evolution, $\langle {\rm AR} \rangle-1$ increases, consistent with the larger number of elongated grains visible in Fig.~\ref{fig_law}d;
external stress further amplify this effect (cf. Figs.~\ref{fig_law}d–f).
{ This demonstrates that GB shear coupling induces the formation of elongated grain shapes, even in the isotropic GB limit. 
This observation offers significant implications for interpreting experimental observation where anisotropic grain morphologies cannot be predicted from the GB energy anisotropy alone~\cite{dillon2010grain,naghibzadeh2024impact}.
}
A related geometric quantity is the average grain roundness $\langle \bigcirc \rangle$; grains with larger aspect ratios naturally exhibit lower roundness.
We plot the degree by which the grain shape deviates from being perfectly circular $\langle \Delta \bigcirc \rangle = 1 - \langle \bigcirc \rangle$  in Fig.~\ref{fig_nonsimilar}c.
Consequently, the ordering of the average deviation from a circle among the four scenarios is
$\langle \Delta\bigcirc_{\rm 12}^{\rm ext}\rangle <\langle \Delta\bigcirc_{\rm 11}^{\rm ext}\rangle <\langle \Delta\bigcirc^{\rm int}\rangle \ll\langle \Delta\bigcirc^{ \gamma}\rangle $.

Our analysis of the various topological and geometric properties of the evolving microstructure shows that internal and applied external stresses drive pronounced deviations from an equiaxed grain morphology, leading instead to more elongated grains and increased inhomogeneity in the distributions of grain size, grain perimeter, and GB length.

\subsection{Internal stresses within grain growth microstructures}

\begin{figure}[b]
\includegraphics[width=0.825\linewidth]{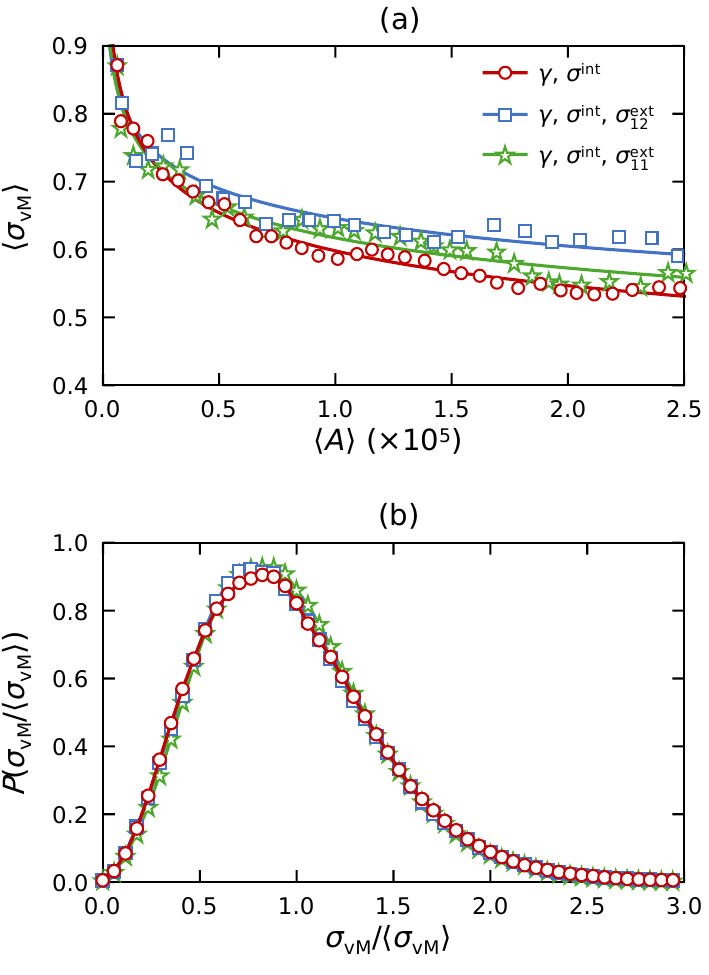}\hspace{-1.78em}%
\caption{The temporal evolution of the internal stress field in the cases of $\gamma,~\sigma^{\rm int}$ (red circles), $\gamma,~\sigma^{\rm int},~\sigma^{\rm ext}_{11}$ (green star) and $\gamma,~\sigma^{\rm int},~\sigma^{\rm ext}_{12}$ (blue squares). (a) Average internal von Mises stress ($\langle\sigma_{\rm vM}\rangle$) vs. mean grain area.
(b) Probability distribution of reduced internal von Mises stress ($\sigma_{\rm vM}/\langle\sigma_{\rm vM}\rangle$).
Note that each probability distribution is averaged over the microstructures with a mean grain area $\langle A \rangle > 5\times 10^4$.
}
\label{fig_vMtime}
\end{figure}

Given that internal stresses strongly modify GG, it is natural to ask: how do internal stresses evolve? 
Figure~\ref{fig_vMtime} shows the temporal evolution of the average internal  shear (von Mises) stress, $\langle \sigma_{\rm vM} \rangle$.
Figure~\ref{fig_vMtime}a, shows that $\langle \sigma_{\rm vM} \rangle$ decrease continuously during GG.
To quantify this behavior, we fit the  internal stress dependence on the mean grain size through   $\langle \sigma_{\rm vM}\rangle = K_\sigma\langle A\rangle^{-n_\sigma}$; the best fit parameters are shown in Table~\ref{tab:fit_parameter} and the fit is very good.
Examination of Fig.~\ref{fig_vMtime}a and the fact that $n_\sigma >0$ show that the stress in the microstructure decays as grains grow; as the density of GBs tends to zero, so does the stress (recall that the GBs are the source of the stress through GB shear coupling). 
It is interesting to observe that, with shear coupling, the average internal stress decays roughly as  $\langle \sigma_{\rm vM}\rangle \sim \langle \mathcal{R} \rangle^{1/4}$ (while suggestive, we do not have a good model for such a relation). 
The relaxation of internal stress is slower in the cases with applied external stress than in those without, following an order opposite to that of the corresponding grain-growth rates (cf. Figs.~\ref{fig_law}a and \ref{fig_vMtime}a).
Figure~\ref{fig_vMtime}b shows the probability distribution of the reduced internal von Mises stress, $P(\sigma_{\rm vM}/\langle \sigma_{\rm vM} \rangle)$.
For cases both with and without applied external stress, this distribution is log-normal-like and remains self-similar throughout GG (see Fig. S3).

\begin{table}[t]\label{tab:fit_parameter}
\centering
\caption{Fitted parameters of the internal stress vs. mean grain area data, $\langle \sigma_{vM} \rangle = K_\sigma t^{n_\sigma}$.}
\renewcommand{\arraystretch}{1.5}
%\begin{ruledtabular}
\small
%\begin{tabular}{ m{4cm}<{\centering} m{2.8cm}<{\centering} m{2.8cm}<{\centering} m{2.8cm}<{\centering} m{3cm}<{\centering}}
\begin{tabular}{ccc}%{p{3cm} p{2.5cm} p{2.5cm}}
\hline\hline
\quad\quad\quad Case \quad\quad\quad & \quad\quad $K_\sigma$ ($\times 10^{5}$) \quad\quad\             & \quad\quad\quad\quad $n_\sigma$ \quad\quad\quad\quad  \\ \hline
$\gamma,~\sigma^{\rm int}$                                      & 0.598 $\pm$ 0.0022                   & 0.129 $\pm$ 0.0030                                     \\ %\hline
$\gamma,~\sigma^{\rm int},~\sigma^{\rm ext}_{\rm 11}$ & 0.617 $\pm$ 0.0035                   & 0.107 $\pm$ 0.0048  \\ %\hline
$\gamma,~\sigma^{\rm int},~\sigma^{\rm ext}_{\rm 12}$   & 0.646 $\pm$ 0.0042                   & 0.094 $\pm$ 0.0054   \\ \hline\hline
\end{tabular}
%\end{ruledtabular}
\end{table}

\begin{figure*}[t]
    \centering
\includegraphics[width=0.99\linewidth]{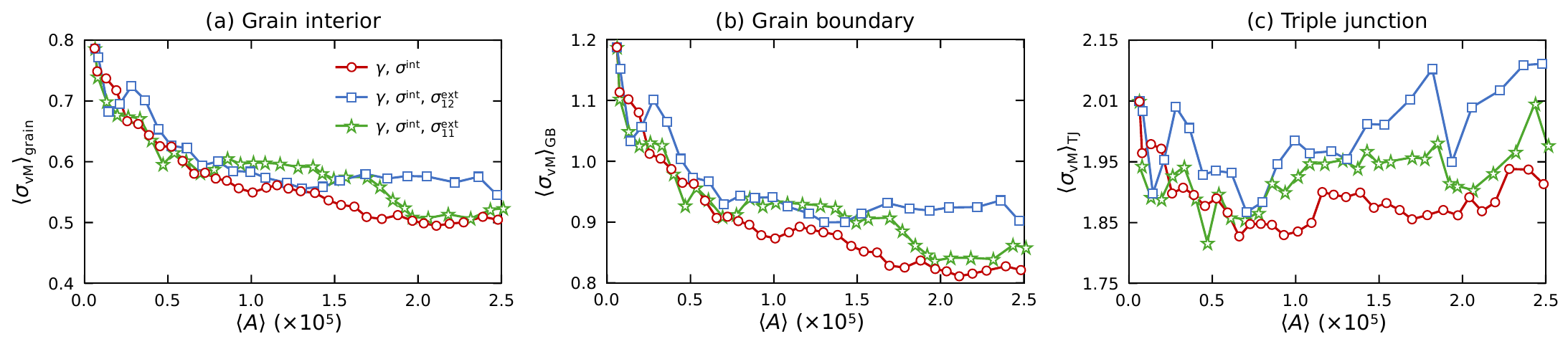}%\hspace{-1.78em}%
\caption{The temporal evolution of the internal stress field (a) in grain interior, (b) at GBs, and (c) at TJs in the cases of $\gamma,~\sigma^{\rm int}$ (red circles), $\gamma,~\sigma^{\rm int},~\sigma^{\rm ext}_{11}$ (green star) and $\gamma,~\sigma^{\rm int},~\sigma^{\rm ext}_{12}$ (blue squares). 
}
\label{fig_vMtopo}
\end{figure*}

\begin{figure*}[t]
\includegraphics[width=0.99\linewidth]{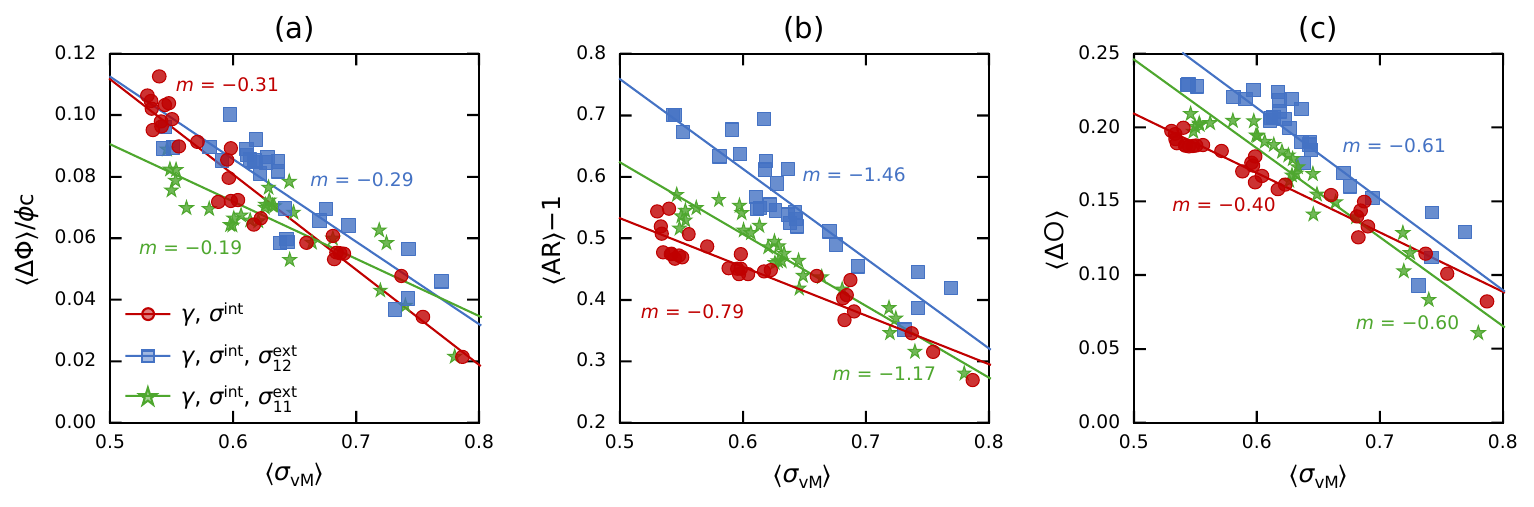}\hspace{-1.78em}%
\caption{
(a) GB degree of faceting $\langle \Delta\Phi \rangle/\phi_c$, (b) grain aspect ratio $\langle {\rm AR} \rangle -1$ and (c) deviation from a circle $\langle\Delta\bigcirc\rangle$ versus the averaged internal von Mises stress.  The cases are $\gamma,~\sigma^{\rm int}$ (red circles), $\gamma,~\sigma^{\rm int},~\sigma^{\rm ext}_{11}$ (green star) and $\gamma,~\sigma^{\rm int},~\sigma^{\rm ext}_{12}$ (blue squares), respectively. Each data point is the microstructure average of a geometric property at a particular mean grain size. The solid lines are linear fits to the data, with slope $m$. Note that time runs from large to small $\langle \sigma_{\rm vM}\rangle$ (right to left).
}
\label{fig_stressgeometry}
\end{figure*}

As shown previously~\cite{qiu2025why}, the internal stress is spatially inhomogeneous, and there is a considerable stress accumulation at GBs and TJs. 
However, how internal stress evolves across different regions of the microstructure remains unclear.
We show the evolution of internal stress within grains, at GBs, and at TJs (as defined by the magnitude of the order parameters in the phase field model) in Fig.~\ref{fig_vMtopo}.
The evolution of the average internal von Mises stress within  grains and at GBs follows the same trend as that measured for the whole microstructure (Fig.~\ref{fig_vMtime}a), i.e., stress relaxation during evolution, with slower decay under applied external stress.
In contrast, the internal stress at TJs is noisy, but is roughly time- or $\langle A \rangle$-independent within the statistics of the simulation (Fig.~\ref{fig_vMtopo}c).
In our continuum model, the origin of internal stress relaxation is the evolution of the disconnection Burgers vector density on GBs~\cite{zhang2017equation,han2022disconnection,sal2022disconnection}, and the stress in the grain interior is the long-range stress originating from shear coupling.
Therefore, it is unsurprising that the internal stress evolution within grains and at GBs follows similar trends, with a relatively smaller stress magnitude and relaxation rate in grain interiors than GBs (due to the $\sim 1/r$ decay of the stress field away from disconnection cores on GBs~\cite{anderson2017}).
However, TJs (with lower mobility than GBs) act as obstacles to GB migration, leading to disconnection pile-up at TJs and increased local internal stress~\cite{thomas2019disconnection,qiu2025why}.
The order of the stress relaxation rates obtained from our PF simulations, 
$\dot{\overline{\langle\sigma_{\rm vM}\rangle}}_{\rm GB} > \dot{\overline{\langle\sigma_{\rm vM}\rangle}}_{\rm grain} \gg \dot{\overline{\langle\sigma_{\rm vM}\rangle}}_{\rm TJ}$, consistent with the argument above.

\section{Discussion}

We demonstrated that internal stress (i) significantly alters the  evolving microstructure and (ii) itself evolves during microstructure evolution. 
How does the evolution of microstructure co-evolve with internal stress?

The co-evolution of stress and the geometric descriptors of microstructure can be described by considering the variation of those descriptors with the averaged internal von Mises stresses, as shown in Fig.~\ref{fig_stressgeometry}.
All of these descriptors (the degree of faceting, grain aspect ratio, and deviation from a circle) decrease linearly with the averaged internal von Mises stress (see the fitted lines). 
While the grain shapes become more isotropic and faceting decreases with increasing average internal von Mises stress, application of an external stress (uniaxial or shear) changes the slopes as $\left|m^{\rm int}\right| < \left|m^{\rm ext}_{\rm 11}\right| < \left|m^{\rm ext}_{\rm 12}\right|$ (the opposite occurs for GB faceting).
External stress strengthens this dependence for the deviation from a circle and aspect ratio, but  weakens it for the degree of faceting.
This is consistent with the effects of external stress on the evolution of the same properties with grain size in Fig.~\ref{fig_nonsimilar}. These results demonstrate that the evolution of the geometric descriptors of the GG microstructure is strongly dependent on internal stress. 
We should note that the average von Mises stress in Fig.~\ref{fig_stressgeometry} increases from left to right, whereas time moves from right to left (i.e., the mean grain size increases and the average von Mises stress decreases with time; see Fig.~S5). 

\begin{figure}[t]
\includegraphics[width=0.9\linewidth]{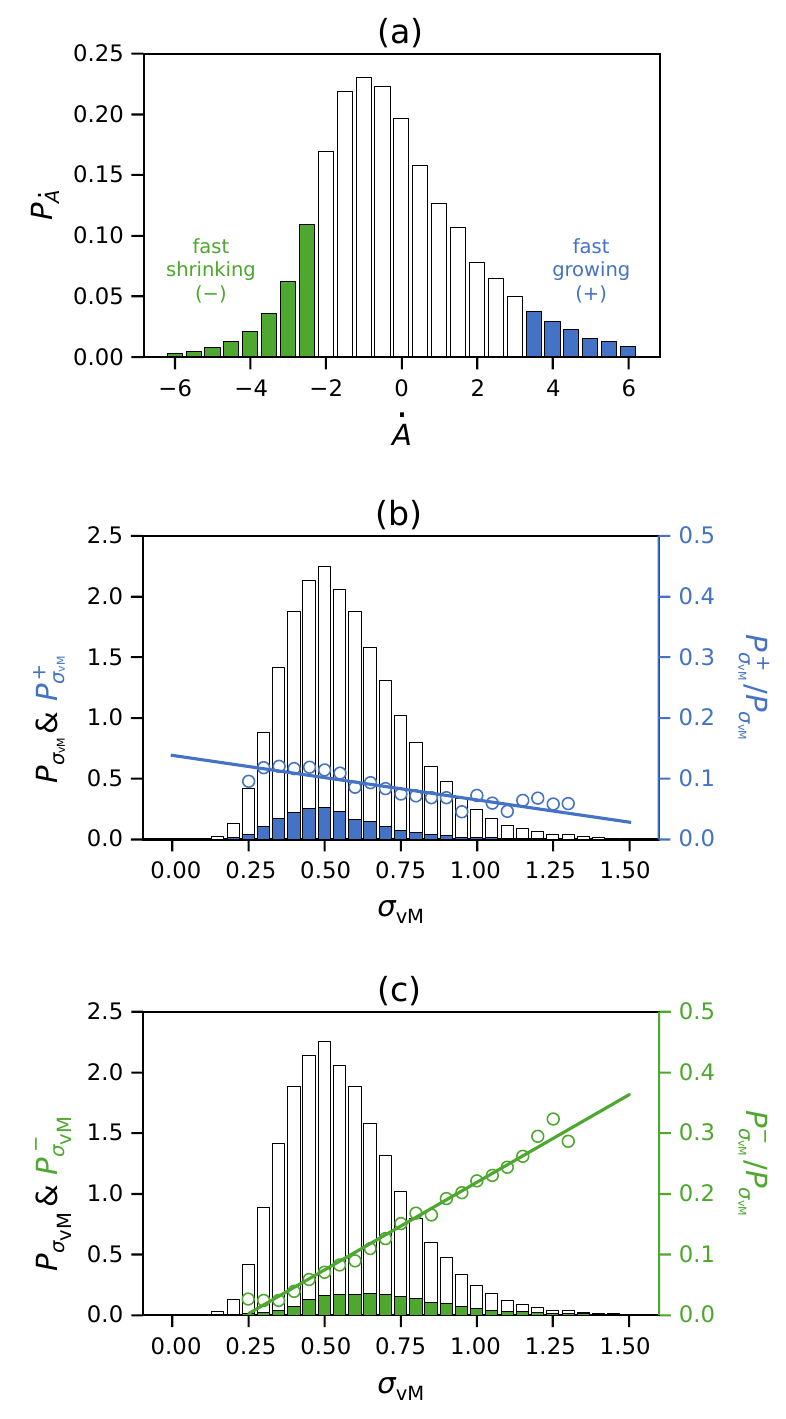}\hspace{-1.78em}%
\caption{(a) Probability distribution of the GG rates in the case of $\gamma,~\sigma^{\rm int}$. The regions shaded in blue and green correspond to the 10\% of the grains that grow the fastest and slowest, respectively.
The probability distributions of internal von Mises stress for all grains (indicated by white bars), (b) for the fast-growing grains (indicated by blue bars), and (c) for the fast-shrinking grains (indicated by green bars). 
The ratios, $P_{\sigma_{\rm vM}}^{\rm +}/P_{\sigma_{\rm vM}}$ and $P_{\sigma_{\rm vM}}^{\rm -}/P_{\sigma_{\rm vM}}$, are plotted as blue and green points.
}
\label{fig_stressgrowth}
\end{figure}

To  gain further insight into the impact of stress on microstructure evolution, we examine the role of internal stress on the growth rate of individual grains. 
Figure~\ref{fig_stressgrowth}a shows the probability distributions of grain-growth rates, analogous to Fig.~\ref{fig_neumann}b. 
We identify the 10\% of the grains that grow the fastest and the 10\% of the grains that shrink the fastest in Fig.~\ref{fig_stressgrowth}a (the fast-growing grains have $\dot{A}>3.5$ are in blue and the fastest shrinking grains have  $\dot{A}<-2.3$ in green). 
The white histogram in Figs.~\ref{fig_stressgrowth}b,c shows the probability distribution of the internal von Mises stress (with internal stress and no external stress applied). 
In Figs.~\ref{fig_stressgrowth}b,c, the blue and green histograms show the parts of the stress distribution corresponding to the fast-growing and fast-shrinking grains, respectively. 
The ratios of the fast-growing and fast-shrinking grains to the total number of grains in each stress bin are plotted as hollow blue and green circles in Figs.~\ref{fig_stressgrowth}b and c, respectively (the best fit to these data is represented by the solid lines).
We find that fast-growing grains tend to be those  with  small internal stress (i.e., the negative relation in Fig.~\ref{fig_stressgrowth}b), whereas fast-shrinking grains tend to have large internal stress (i.e., the positive relation in Fig.~\ref{fig_stressgrowth}c).

To confirm that the inclusion of internal stress as a \emph{driving force} in the microstructure evolution is the cause of the correlation between internal stress and GG rate in Figs.~\ref{fig_stressgrowth}b,c, we performed a similar analysis for microstructure evolution only driven by capillarity (curvature flow; see Figs.~S6 in SM); in other words,  the internal stress is computed but not included in the driving force for GB migration.  
These results show that a correlation between internal stress and growth rate exists only when internal stress is included as a driving force for GB migration.

\begin{figure}[t]
\includegraphics[width=0.99\linewidth]{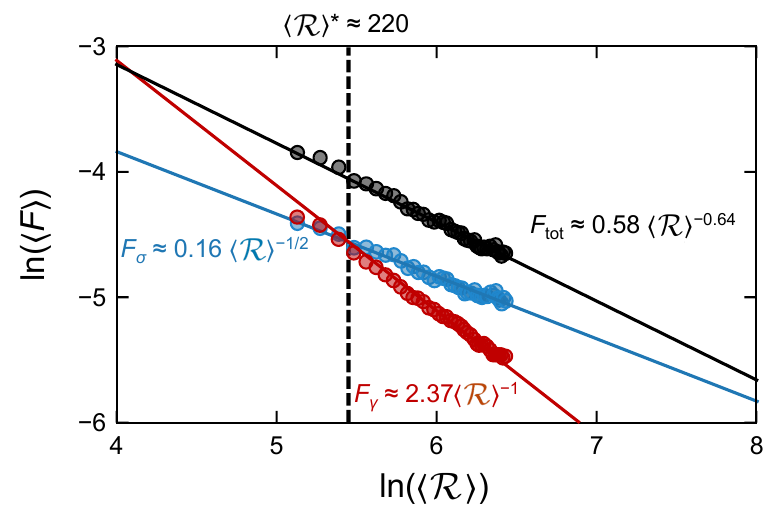}\hspace{-1.78em}%
\caption{Evolution of the average capillarity driving force (red circles),  elastic driving force (blue circles) and total driving force (black circles) vs. mean grain size $\langle\mathcal{R}\rangle$ during GG. 
The red solid line is the fit to the classical capillarity law, $F\sim 1/\mathcal{R}$ (the fitted exponent is $-0.90$) and the blue solid line is $F\sim 1/\sqrt{\mathcal{R}}$  (the fitted exponent is $-0.51$).
The black solid line is best fitted as $F\sim 1/\mathcal{R}^{-0.64}$.
}
\label{fig_F}
\end{figure}

In Section~\ref{sec:law}, Fig.~\ref{fig_law}a and Table~\ref{tab:fit_parameter_size}, we found  that the GG exponent is larger than 1 when internal stresses are considered (in curvature flow, the exponent is 1). 
To understand the origin of this phenomenon, we track the evolution of the average capillarity force ($F_\gamma$) and elastic driving force ($F_\sigma$) along all GB segments as a function of grain size during GG; see the red and blue circles in Fig.~\ref{fig_F}.    
We fit these data to the function $F \sim \langle \mathcal{R}\rangle^{-n_F}$ (solid red and blue lines in the figure).
According to conventional curvature-driven GG theory, the capillarity force scales inversely with the mean grain size (i.e., $F_\gamma \sim \langle \mathcal{R} \rangle^{-1}$). 
Our simulations yield a fitted exponent of $n_F \approx -0.90$ for $F_\gamma$, which is close to the theoretical value of $-1$ (the small deviation can be attributed to the diffuse-interface numerical method and the lack of data at very large size).
In contrast, $n_F = -0.51\approx -1/2$ for the elastic driving force $F_\sigma$ data. 
This difference in scaling implies that while capillarity dominates the early stages of growth when grains are small, elastic driving forces  dominate GG kinetics at large grain sizes  {(this can also be seen by comparing the evolution of the probability distributions of the capillarity and elastic driving forces in Figs.~S7a,b).}
The exponent $n_F \approx -0.5$ of the elastic driving force can be understood in terms of  disconnection pile-ups at the triple junction, similar to the yield stress decaying as $\langle \mathcal{R}\rangle^{-0.5}$ when the bulk dislocations pile up at GBs in Hall-Petch relation~\cite{hall1951deformation} (see the theoretical model developed in SM). 
The crossover point is around $\langle \mathcal{R} \rangle \approx 220$.
If we fit the total force on GBs as a function of grain size $\langle \mathcal{R} \rangle$, we find $n_F = -0.64$, 
which is between these two limits. 

If we assume $\rmd \langle \mathcal{R} \rangle /\rmd t\sim F$~\cite{burke1952recrystallization}, the GG law follows $\langle A\rangle = \langle \mathcal{R}\rangle^2 \propto t$ if the driving force is only $F_\gamma$ and 
$\langle A\rangle = \langle \mathcal{R}\rangle^2 \propto t^{4/3}$ if only $F_\sigma$. Our numerical results for $F_\textrm{tot}$ versus $\rmd \langle \mathcal{R} \rangle$ suggest that $n_A\approx1.22$, which is $n_A\in [1, 4/3]$ when both capillarity force and elastic driving force are considered.
Our fitted $n_A$ results in Table~\ref{tab:fit_parameter_size} perfectly fall in this range and are consistent with the fitted force data in Fig.~\ref{fig_F}.

We note that the GG exponent measured in experiments is rarely larger than 1.
The discrepancy between our simulation/theoretical values of $n_A$ and those from experiments may stem from other forces present in experiments and not included in our model (e.g., solute drag) or from simplifications in the current continuum model. 
In our phase-field and disconnection-based approaches~\cite{han2022disconnection,sal2022disconnection,qiu2025why}, only a single disconnection mode is considered, whereas in real GB migration, multiple disconnection modes~\cite{tian2024grain,gautier2025quantifying} may play a role. 
This limitation may lead to an overestimation of the shear coupling factor and the resulting internal stress field~\cite{gautier2021shear,gautier2025quantifying}. 
The inclusion of multiple disconnection modes will lower the elastic driving force (moving the blue line downward in Fig.~\ref{fig_F}), shifting the critical mean grain size where the capillarity and elastic driving forces cross over to a  larger mean grain size (where other effects, such as solute drag, may dominate). 
The current model also neglects other possible stress relaxation pathways such as annealing twin formation~\cite{fullman1951formation,thomas2016twins} and dislocation emission from grain boundaries~\cite{li2006mechanical,bizana2023kinetics}.

The coupling of stress and GB motion has been clearly demonstrated by experiments on thin, quasi-two-dimensional polycrystals with nanoscale average grain sizes~\cite{tian2024grain,gautier2025quantifying}.  
The results in Fig.~\ref{fig_size} indicate that the influence of shear coupling increases with grain size, and this is consistent with reports that GB migration in three-dimensional polycrystals with micron-scale grain sizes is inconsistent with curvature flow~\cite{zhang2020grain,xu2024grain,xu2026can}.  
However, we note that the three-dimensional X-ray data provide no evidence of internal strain from defects or grain rotations in nanocrystalline samples that are above the noise floor ($\sim0.1$°  lattice rotation)~\cite{bhattacharya2021grain}.  
On the other hand, observations in thin films~\cite{tian2024grain} show  rotations larger than this and should be easily detectable in  X-ray measurements.  
If the strains were localized to regions very close to  grain boundaries (i.e., below the spatial resolution of X-ray microscopy), they could evade detection (although such strain distributions are not typically found in higher-resolution studies of GBs).  
Another possibility is that the dimensionality of the experiment influences the results.  
Simulations of shear-coupled GB migration in one dimension have illustrated the importance of free and fixed boundary conditions~\cite{thomas2017reconciling}.  
In the fixed-boundary-condition case, multiple disconnection modes are required to sustain GB migration.  
Quasi-two-dimensional samples have free boundary conditions at the top and bottom of the film, and Gauthier et al.~\cite{gautier2025quantifying} have shown that vertical displacements, normal to the film plane, occur when boundaries in the film migrate.  
Such displacements would not be accommodated in the three-dimensional case.  
Additional experiments and simulations will be required to understand the effect of dimensionality on shear-coupled motion in polycrystals.

\begin{comment}
However, we note that experimental GG is rarely observed to continuously accelerate, and experimentally measured exponents $n_A$ are typically below 1~\cite{burke1949some,bolling1958grain,hu1974grain,holm2010grain}.
The discrepancy between our simulation/theoretical values of $n_A$ and those from experiments likely stems from simplifications in the current continuum model. 
In our phase-field and disconnection-based approaches~\cite{han2022disconnection,sal2022disconnection,qiu2025why}, only a single disconnection mode is considered, whereas real GB migration involves multiple disconnection modes~\cite{tian2024grain,gautier2025quantifying}. 
This limitation may lead to an overestimation of the shear coupling factor and the resulting internal stress field~\cite{gautier2021shear,gautier2025quantifying}.
Moreover, the inclusion of multiple disconnection modes could increase the exponent $n_F$, potentially eliminating the predicted acceleration or even leading to slower growth over time—consistent with experimental trends. 
Additional stress relaxation pathways are not included in our model, such as annealing twin formation~\cite{fullman1951formation,thomas2016twins} and dislocation emission from grain boundaries~\cite{li2006mechanical,bizana2023kinetics}, may further contribute to this discrepancy. 
\end{comment}

%\clearpage

%%%%%%%%%%%%%%%%%%%%%%%%%%%%%%%%%%%%%%%%%%%%%%%%%%%%%%%%%%%%%%%%%%%
\section{Conclusions}
We conducted a series of large-scale phase-field simulations to investigate and quantify the effects of internal and external stresses on GG, using a continuum model that reconciles curvature-driven and shear-coupled (disconnection-controlled) GB migration.
We investigated (i) the evolution of the geometric and topological properties of the microstructures, (ii) the evolution of the internal stress field, and  (iii) analyzed how these  are correlated and can be reconciled with each other.

Our main conclusions are as follows.
\begin{enumerate}
    \item Compared to microstructure evolution by mean-curvature flow alone, internal stress leads to less homogeneous and less equiaxed grains, enhanced GB faceting, increased grain aspect ratio, reduced grain roundness, and a shift of the grain-size distribution to smaller size, in agreement with experimental and atomistic results.
\item The average internal stress gradually relaxes during GG, with stress accumulation occurring around some TJs.
%\item Internal stress between neighboring grains is strongly correlated at small mean grain sizes, while this correlation decreases as grains grow.
\item Grains with smaller internal stress tend to grow faster, whereas those with larger internal stress shrink faster.
\item Internal stress generation associated with GB shear coupling tends to accelerate GG (i.e., produce larger GG exponents).
\item The application of external stress increases the grain aspect ratio, weakens GB faceting and grain roundness, and reduces the overall degree of internal-stress relaxation over time.

\end{enumerate}

Our results demonstrate that incorporating internal stress  captures fundamental features of GG, as observed in experiments and atomistic simulations, which curvature flow alone fails to reproduce. 
The formulation underlying our model, grounded in disconnection-mediated GB migration, enables simulations of detailed microstructural features that result from both intrinsic driving forces in GG - curvature flow AND stress generation.  The  parameterization of our model, although minimal, is theoretically applicable to all FCC metals~\cite{qiu2025why}, and can be readily extended to other crystalline metals and even to ceramics by adjusting material-specific parameters.

\section*{Acknowledgements}
CQ and DJS acknowledge support from the Hong Kong Research Grants Council Collaborative Research Fund C1005-19G and General Research Fund 17210723 and 17200424. JH acknowledges support of the National Key R \& D Program of China 2021YFA1200202 and the Early Career Scheme (ECS) grant from the Hong Kong Research Grants Council CityU21213921. GSR acknowledges support by the National Science Foundation under DMREF Grant No. 2118945. MS acknowledges the support of the Deutsche Forschungsgemeinschaft (DFG, German Research Foundation) project No. 447241406 and 570666382.

%\section*{Declaration of competing interest}
%The author Gregory S. Rohrer is Coordinating Editor for Acta Materialia and was not involved in the editorial review or the decision to publish this article.

%apsrev4-2.bst 2019-01-14 (MD) hand-edited version of apsrev4-1.bst
%Control: key (0)
%Control: author (8) initials jnrlst
%Control: editor formatted (1) identically to author
%Control: production of article title (0) allowed
%Control: page (0) single
%Control: year (1) truncated
%Control: production of eprint (0) enabled
%

\clearpage
\newpage

\onecolumngrid

\begin{center}
  \textbf{\large \hspace{5pt} SUPPLEMENTARY INFORMATION \\ \vspace{0.2cm} Shear-Coupled Grain Growth Statistics \\}
\vspace{0.4cm}   Caihao Qiu,$^{1,2}$, David J. Srolovitz,$^{2,3}$, Gregory S. Rohrer$^{4}$, Jian Han$^{1}$, Marco Salvalaglio$^{5,6}$ \\[.1cm]
  {\itshape \small
  ${}^1$Department of Materials Science and Engineering,\\ City University of Hong Kong,  Hong Kong SAR, China
  \\ 
  ${}^2$Department of Mechanical Engineering, The University of Hong Kong, Pokfulam Road, Hong Kong SAR, China
 \\
  ${}^3$Materials Innovation Institute for Life Sciences and Energy (MILES), The University of Hong Kong, Shenzhen, China
 \\
  ${}^4$Department of Materials Science and Engineering, Carnegie Mellon University, Pittsburgh, PA, USA
   \\
  ${}^5$Institute of Scientific Computing, TU Dresden, 01062 Dresden, Germany
   \\
  ${}^6$Dresden Center for Computational Materials Science, TU Dresden, 01062 Dresden, Germany
 }
  %\vspace{0.1cm}{\small
  % ${}^*$marco.salvalaglio@tu-dresden.de\\
  % ${}^\dagger$monica.bollani@ifn.cnr.it\\
  % ${}^\ddagger$marco.abbarchi@im2np.fr}
%(Dated: \today)\\
\vspace{0.5cm}
\end{center}

\twocolumngrid

\setcounter{equation}{0}
\setcounter{figure}{0}
\setcounter{table}{0}
\setcounter{section}{0}
\setcounter{page}{1}

\renewcommand{\thesection}{S-\Roman{section}}
\renewcommand{\theequation}{S-\arabic{equation}}
\renewcommand{\thefigure}{S-\arabic{figure}}
\renewcommand{\bibnumfmt}[1]{[S#1]}
\renewcommand{\citenumfont}[1]{S#1}

\section{Grain growth exponent}
In the main text, we find that the exponent of the elastic driving force on GBs vs. mean grain size is around $-0.5$, leading to a grain-growth exponent of 4/3 for a purely elastic driving force.
Here, we propose a model to explain this phenomenon.

As widely accepted, GB migration is mediated by disconnection motion~\cite{han2018grainSI}.
When a pair of disconnections is nucleated in the middle of the GB and glides away from each other, the elastic energy caused by disconnection gliding can be dissipated rapidly (this is an assumption built into our continuum model)~\cite{han2022disconnectionSI,sal2022disconnectionSI}.
%Then, why does the total elastic energy dissipate so slowly in our current simulations?
However, the pair of disconnections will be blocked by the two ends of the GB, i.e., triple junctions (TJs);
The elastic energy caused by the disconnection pileup at TJs is the main cause of the \emph{slow dissipation of the total elastic energy}~\cite{thomas2019disconnectionSI}.

\begin{figure}[t]
\includegraphics[width=0.99\linewidth]{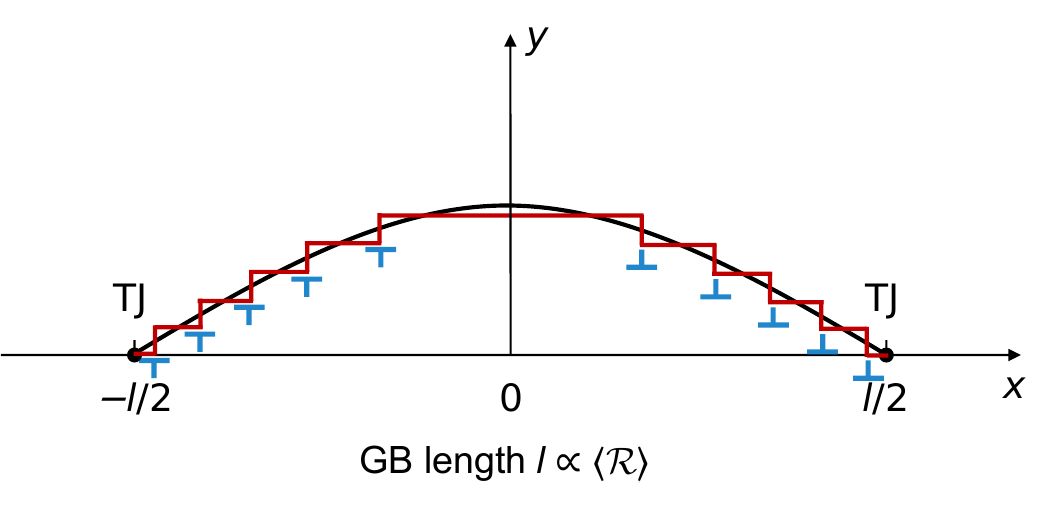}\hspace{-1.78em}%
\caption{Illustration of disconnections (red steps with blue dislocation characters) pileup at TJs located at $(\pm l/2,~0)$ on the GB (black line). 
}
\label{fig_B}
\end{figure}

Figure~\ref{fig_B} provides an illustration of a GB terminated at two TJs separated by a distance of $l$.
This is similar to the case of lattice dislocation pile-up at GBs (i.e., the Hall-Petch effect).
Assuming that the two TJs are immobile, the nucleated disconnections due to the applied shear stress (and/or the elastic interactions among all GB disconnections) will glide toward and be hindered at TJs.
The GB will reach an equilibrium profile with the analytical solution for the equilibrium Burgers vector density distribution (again, similar to the classical double-ended pileup of lattice dislocations) under an applied shear stress $\tau$ is~\cite{anderson2017SI,han2018grainSI}
\begin{equation}
    \rho^b(x) = \frac{2(1-\nu)\tau}{\mu b} \frac{2x/l}{\sqrt{1-(2x/l)^2}},
\end{equation}
where $\nu$ and $\mu$ are Poisson's ratio and shear modulus,
and the equilibrium GB profile is~\cite{han2018grain}
\begin{equation}
    \begin{aligned}
    y(x) &=\int_{-l/2}^{l/2}\rho^h(x^\prime)\rmd x^\prime = -\frac{1}{\beta} \int_{-l/2}^{l/2} \rho^b (x^\prime) \rmd x^\prime \\& = \frac{(1-\nu)\tau l}{\beta \mu b }\sqrt{1-(2x/l)^2},
    \end{aligned}
\end{equation}
where $h$ is disconnection step height, $\rho^h = -\beta\rho^b$ is the step density, $\beta=b/h$ is the shear coupling factor of the GB disconnection.
In the case of grain growth and GB disconnection, $\tau$ is related to the internal (and externally applied) shear stress on the GB.

The total work/energy due to this kind of pileup is
\begin{equation}
    \begin{aligned}
    W &= \frac{\tau b}{2}\int_{-l/2}^{l/2} x \rho^b(x) {\rm d} x \\ &= \frac{(1-\nu)\tau^2}{\mu} \int_{-l/2}^{l/2}  \frac{2x^2/l}{\sqrt{1-(2x/l)^2}} {\rm d} x \\ &= \frac{\pi(1-\nu)l^2\tau^2}{8\mu}.
    \end{aligned}
\end{equation}
The related driving force is
\begin{equation}
    f = -\frac{\partial W}{\partial{l}} = -\frac{\pi(1-\nu)l\tau^2}{4\mu}.
\end{equation}

When the shear stress $\tau$ (i.e., shear stress to drive disconnection motion) exceeds a certain critical value $f_{\rm c}$, the piled-up disconnections can be driven to transmit into other two GBs meeting at the TJ or into the surrounding grains, leading to \textbf{TJ motion (and the consequent GB migration)} and \textbf{the relaxation of the stored elastic energy}.
The critical value of $\tau$ is 
\begin{equation}
    \tau_{\rm c} = \sqrt{\frac{4\mu f_{\rm c}}{\pi(1-\nu)}} l^{-0.5}.
\end{equation}
Since the TJ separation length (and the GB length) is proportional to the mean grain size, the critical shear stress is 
\begin{equation}
    \tau_{\rm c} = C \langle D \rangle^{-0.5}.
\end{equation}
The elastic driving force for GB migration is $\boldsymbol{\tau}\cdot\boldsymbol{\beta}$, where $\boldsymbol{\tau}$ is the (resolved) shear stress $\tau_{\rm c}$ on GB derived above and $\boldsymbol{\beta}$ is the constant GB shear coupling factor.
Hence, the analytical exponent of the elastic driving force vs. the mean grain size is $n^{\sigma}_F = -$0.5, consistent with the value obtained from our simulations ($n_F^{\rm \sigma} = -0.51$).

\section{Supplementary Results: Temporal evolution of the probability distributions of the geometric properties and internal stress}

\begin{figure*}[h]
\includegraphics[width=0.99\linewidth]{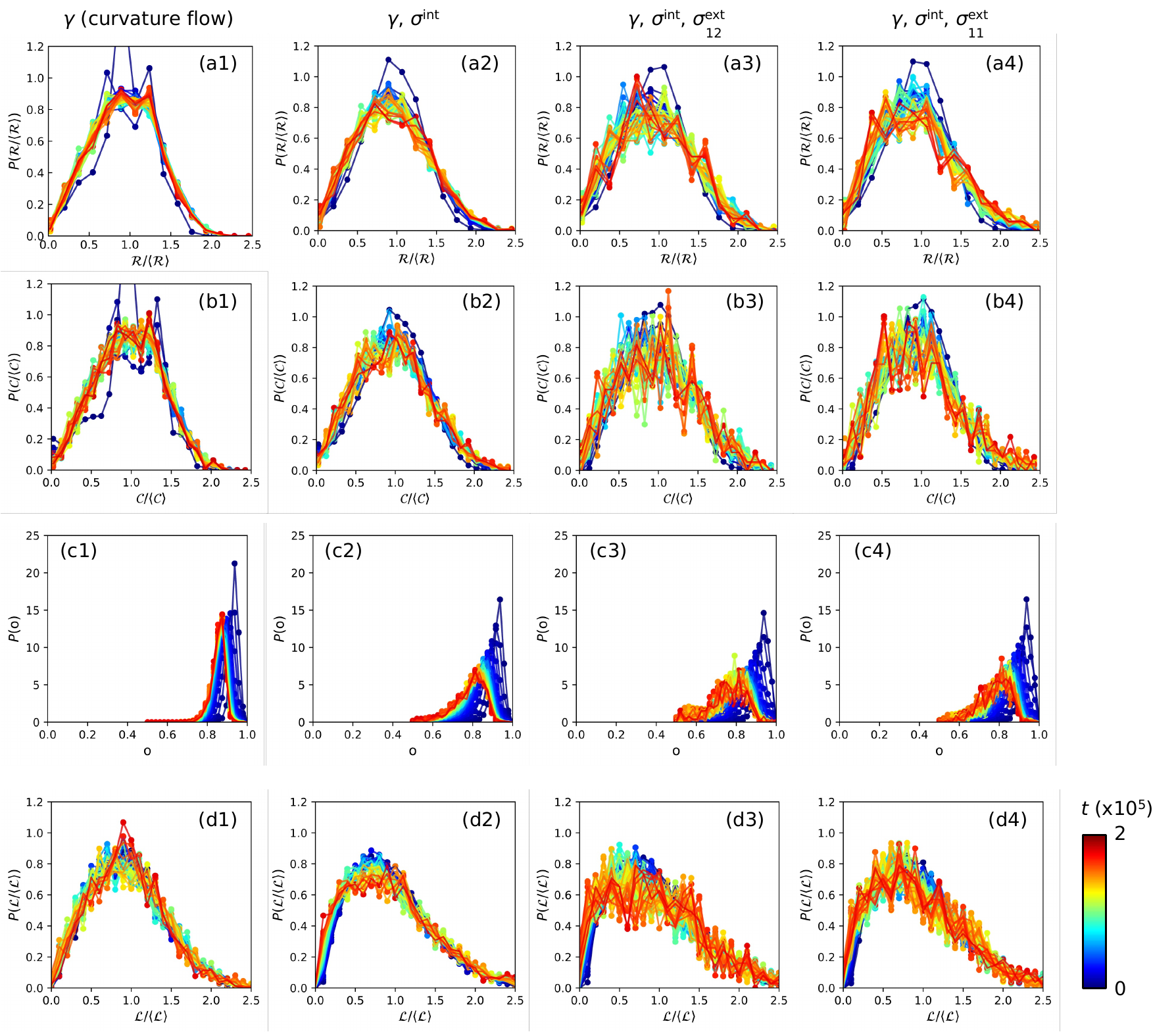}\hspace{-1.78em}%
\caption{Temporal probability distributions of (a1)-(a4) reduced grain size $\mathcal{R}/\langle \mathcal{R} \rangle$, (b1)-(b4) reduced grain perimeter $\mathcal{C}/\langle \mathcal{C} \rangle$, (c1)-(c4) grain roundness $\bigcirc$ and (d1)-(d4) reduced GB length $\mathcal{L}/\langle \mathcal{L} \rangle$. 
The four columns refer to the PF results for the case of curvature flow with internal stress, external shear stress, and external tensile stress, respectively.
Lines in blue are the initial results, and lines in red are the results at the time of $2\times10^5$.
}
\label{fig_geo}
\end{figure*}

In the main text, we investigate the geometric properties of microstructures by analyzing their time-averaged probability distributions.
A question remains as to whether the temporal evolution of the geometric properties can reach a ``steady state'' such that time averaging is meaningful.
Figure~\ref{fig_geo} shows the temporal probability distributions of reduced grain size $\mathcal{R}/\langle \mathcal{R}\rangle$, reduced grain perimeter $\mathcal{C}/\langle \mathcal{C}\rangle$, grain roundness $\langle \bigcirc \rangle$ and reduced grain boundary (GB) length $\mathcal{L}/\langle \mathcal{L}\rangle$.
In each subplot, the line colors represent the simulation time, progressing from blue (initial results) to red (by $2\times10^5$).
The four columns correspond to the cases of curvature flow ($\gamma$), with internal stress ($\gamma,\sigma^{\rm int}$), with external shear stress ($\gamma,\sigma^{\rm int},\sigma_{12}^{\rm ext}$), and with external tensile stress ($\gamma,\sigma^{\rm int},\sigma_{11}^{\rm ext}$), respectively.
The commonality is that the initial probability distribution differs from the results at later times, and they all tend to converge to a similar profile as time evolves.
The special case is the grain roundness, defined as $\bigcirc = 4\pi A/\mathcal{C}^2=4\pi \mathcal{R}^2/\mathcal{C}^2$ shown in Figs.~\ref{fig_geo}(c1)-(c4).
The convergence to a similar distribution takes much longer than the reduction in grain size and grain perimeter.

\begin{figure*}[h]
\includegraphics[width=0.99\linewidth]{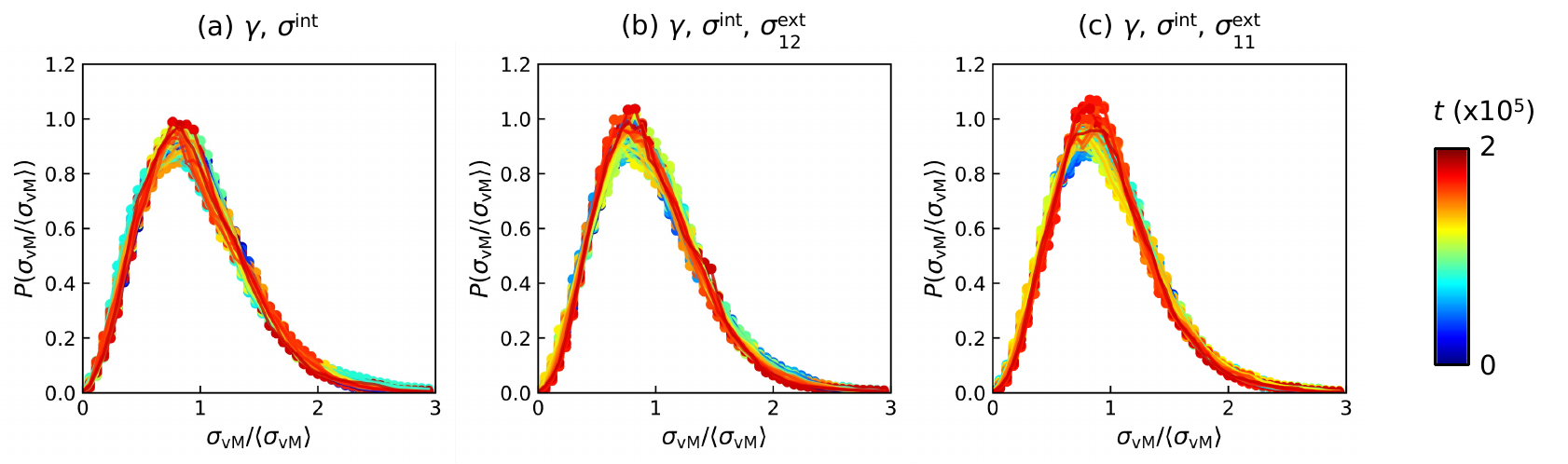}\hspace{-1.78em}%
\caption{Temporal probability distributions of reduced internal von Mises stress $\sigma_{\rm vM}/\langle \sigma_{\rm vM} \rangle$ in the cases with (a) only internal stress, (b) internal stress and applied external shear stress, and (c) internal stress and applied external tensile stress. 
}
\label{fig_str}
\end{figure*}

As for the evolution of internal stresses, we demonstrate that temporal probability distributions of reduced internal von Mises stress $\sigma_{\rm vM}/\langle \sigma_{\rm vM}\rangle$ are always self-similar during grain growth, see Fig.~\ref{fig_str}.
Notably, this occurs without an obvious initial far-from-equilibrium distribution seen in the geometric properties, indicating a fundamentally different relaxation pathway.

\begin{figure*}[h]
\includegraphics[width=0.99\linewidth]{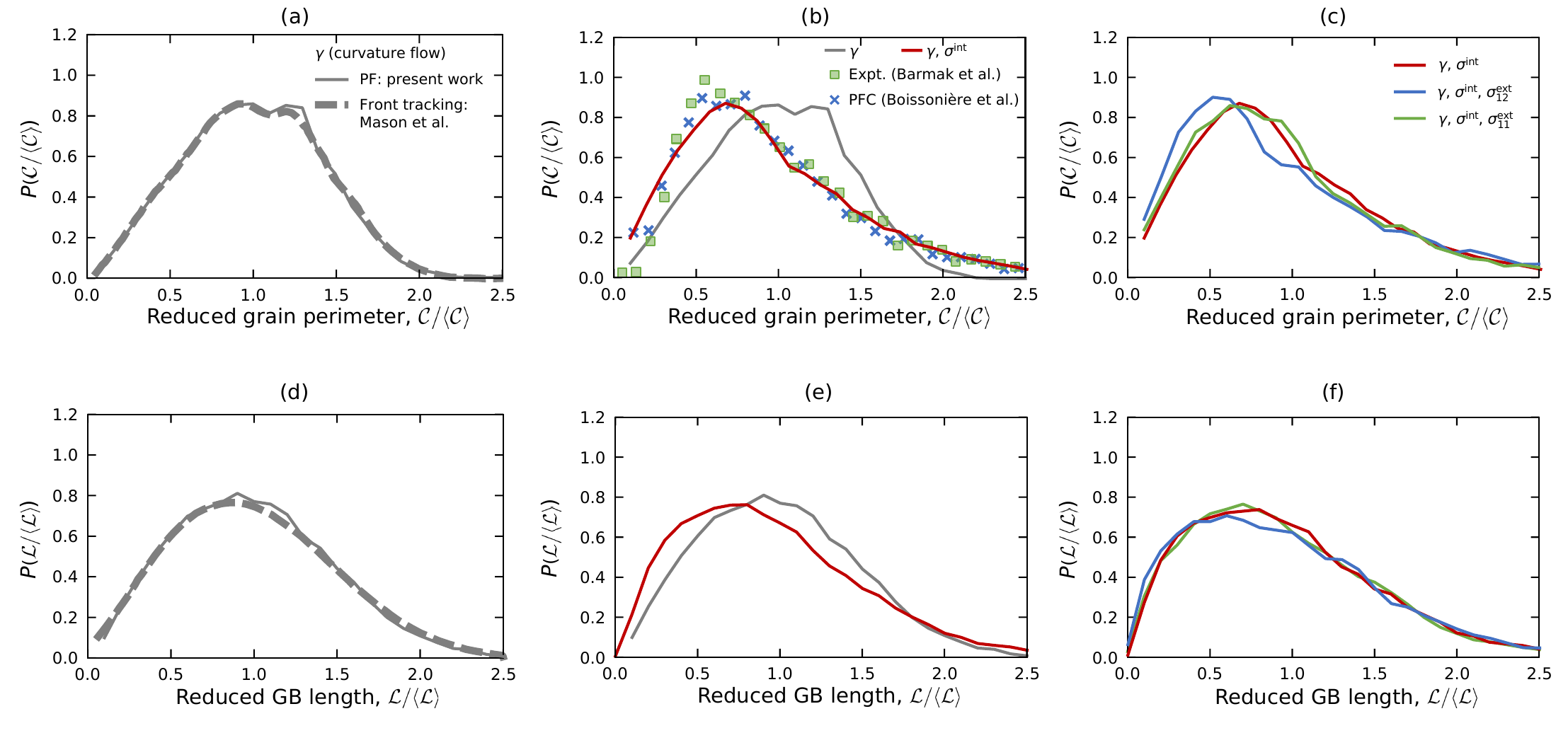}\hspace{-1.78em}%
\caption{Geometric properties of microstructures. probability distributions of (a)-(c) reduced grain perimeter $\mathcal{C}/\langle \mathcal{C} \rangle$, (g)-(i) reduced GB length $\mathcal{L}/\langle \mathcal{L} \rangle$. 
The grey, red, green, and blue lines correspond to our PF results for the case of curvature flow with internal stress, external tensile stress, and external shear stress, respectively.
The thick grey dotted lines in (a), (d), (g), (j) are the results obtained by the  front-tracking method~\cite{mason2015geometricSI}.
Green squares and blue crosses are the results from experiments in an Al thin film~\cite{barmak2013grainSI} and PFC simulations~\cite{la2019statisticsSI}.}
\label{fig_geosm}
\end{figure*}

\begin{figure}[t]
\includegraphics[width=0.99\linewidth]{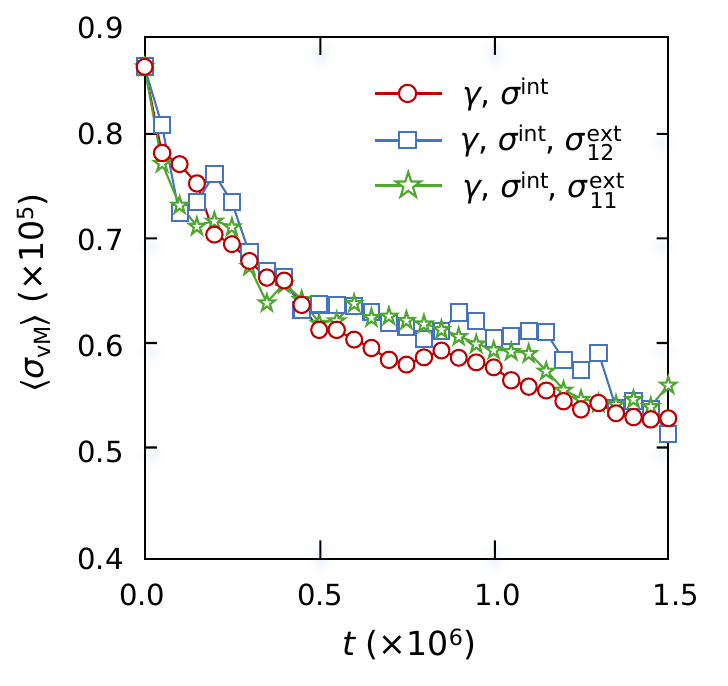}\hspace{-1.78em}%
\caption{Temporal evolution of averaged internal von Mises stress versus time in the cases of $\gamma,~\sigma^{\rm int}$ (red circles); $\gamma,~\sigma^{\rm int},~\sigma^{\rm ext}_{11}$ (green star); and $\gamma,~\sigma^{\rm int},~\sigma^{\rm ext}_{12}$ (blue squares)..
}
\label{fig_stresst}
\end{figure}

\begin{figure}[h]
\includegraphics[width=0.99\linewidth]{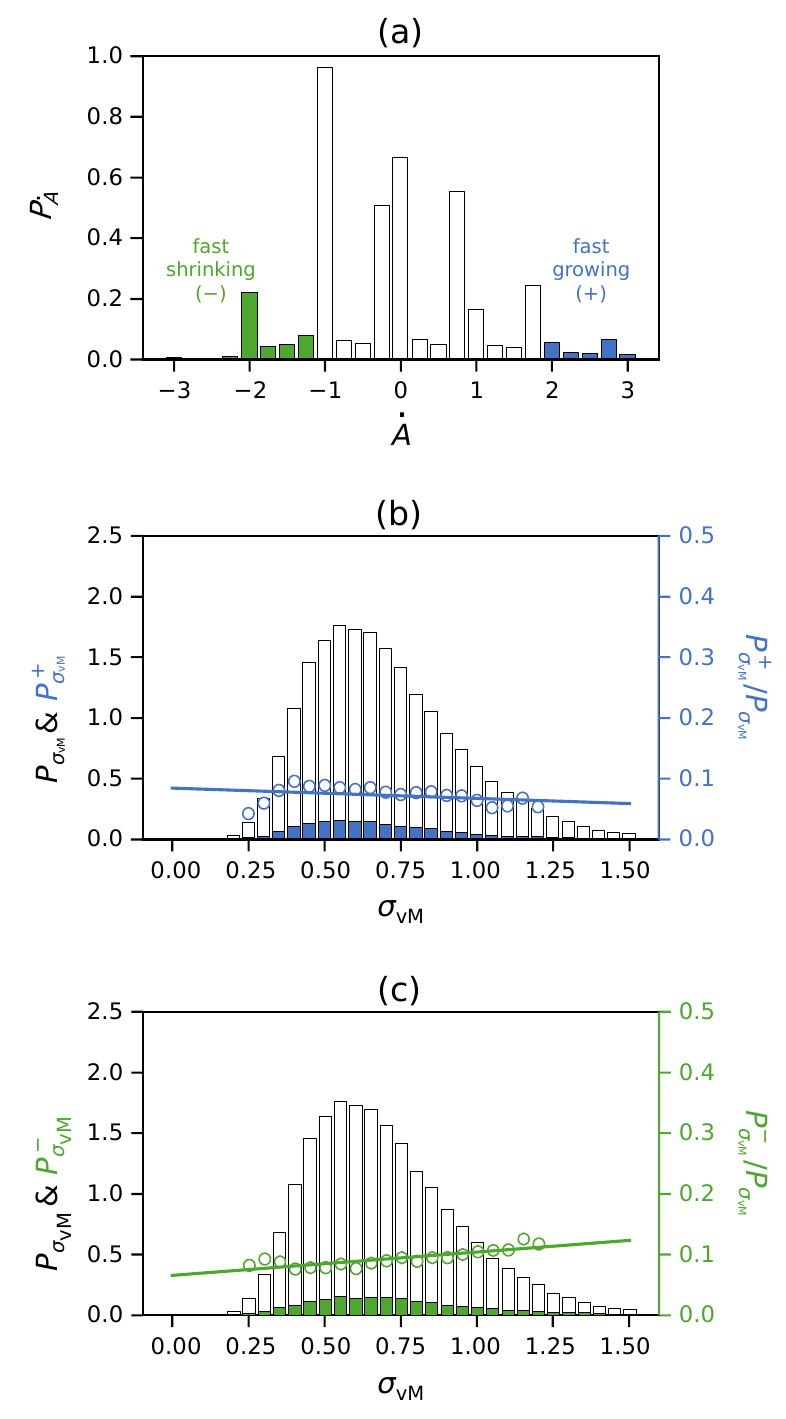}\hspace{-1.78em}%
\caption{(a) Probability distribution of the grain growth rates in the case of mean curvature flow $\gamma$.
The probability distributions of ``virtual'' internal von Mises stress for all grains (indicated by white bars), (b) for the fast-growing grains (indicated by blue bars), and (c) for the fast-shrinking grains (indicated by green bars). 
The ratios, $P_{\sigma_{\rm vM}}^{\rm fast~grow}/P_{\sigma_{\rm vM}}$ and $P_{\sigma_{\rm vM}}^{\rm fast~shrink}/P_{\sigma_{\rm vM}}$, are plotted as blue and green points.}
\label{fig_growthh}
\end{figure}

\begin{figure*}[h]
\includegraphics[width=0.99\linewidth]{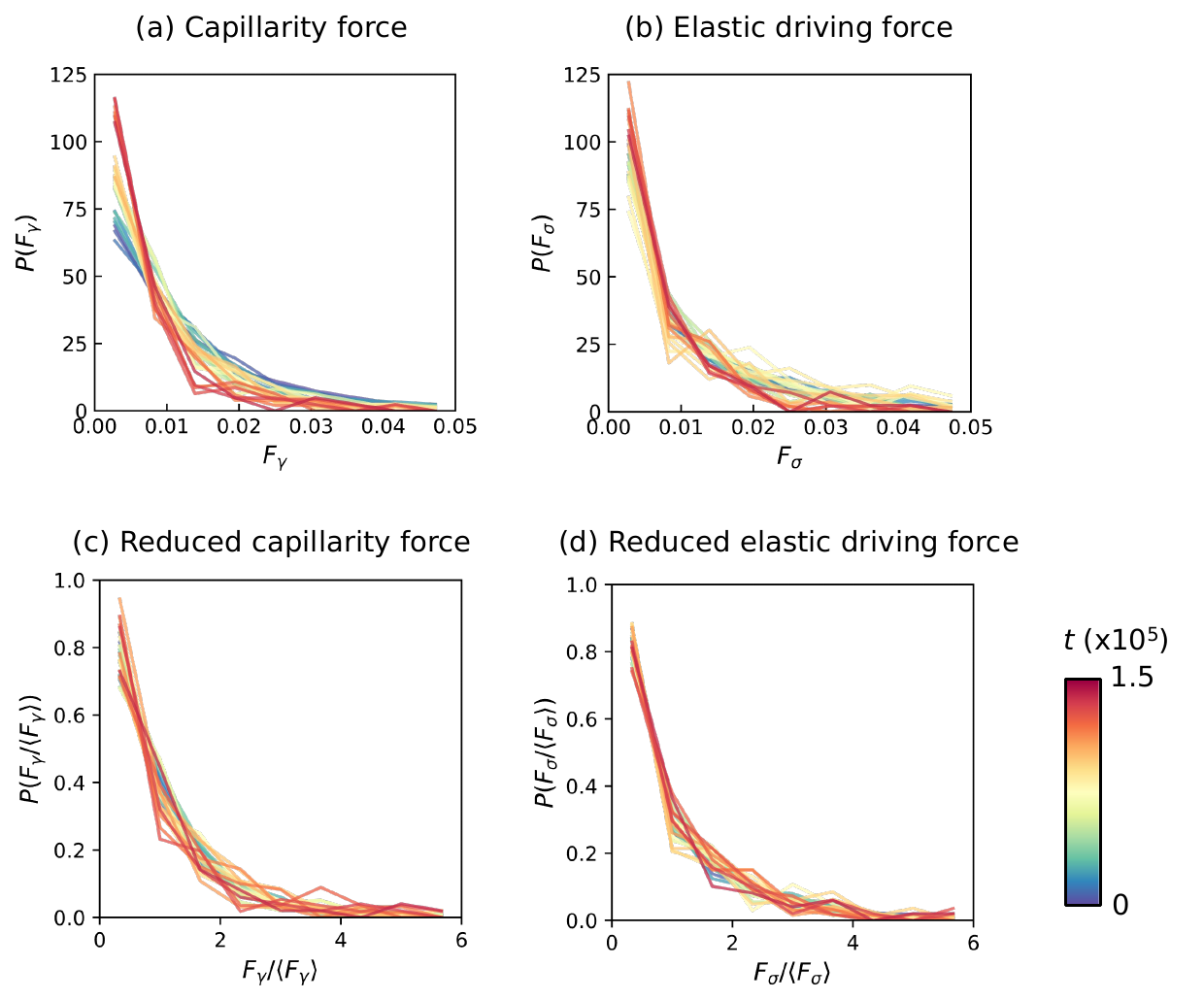}\hspace{-1.78em}%
\caption{Temporal probability distributions of (a) the capillarity force, (b) the elastic driving force, 
(c) reduced capillarity force and (d) reduced elastic driving force.
Lines in blue are the initial results, and lines in red are the results at the time of $1.5\times10^5$.
}
\label{fig_F_dis}
\end{figure*}

\end{document}